\UseRawInputEncoding
\documentclass[aps,onecolumn,10pt]{revtex4}
\usepackage{amsmath}
\usepackage{amssymb}
\usepackage{graphicx}
\usepackage{hyperref}
\usepackage{float}
\usepackage{slashed}
\hypersetup{colorlinks=true,linkcolor=red,citecolor=green,urlcolor=blue}
\numberwithin{equation}{section}
\numberwithin{equation}{section}

\begin{document}
\allowdisplaybreaks
\setcounter{equation}{0}

\title{Structure of conformal gravity in the presence of a scale breaking scalar field}

\author{Philip D. Mannheim}
\affiliation{Department of Physics, University of Connecticut, Storrs, CT 06269, USA \\
 philip.mannheim@uconn.edu\\ }

\date{January 17 2022}

\begin{abstract}

We revisit the structure of conformal gravity in the presence of a c-number, conformally coupled,  long range, macroscopic scalar field. And in the static, spherically symmetric case discuss two classes of exact exterior solutions. In one solution the scalar field has a constant value and in the other solution, which is due to Brihaye and Verbin, it has a dependence on the radial coordinate, with the two exterior solutions being relatable by a conformal transformation. In light of these two solutions Horne and then Hobson and Lasenby raised the concern that the fitting of conformal gravity to galactic rotation curves had been misapplied and thus called the successful fitting of the conformal theory into question. In this paper we show that the analysis of Brihaye and Verbin needs to reappraised, with their reported result not being as general as they had indicated, but nonetheless being valid in the particular case that they studied. For the analyses of Horne and of Hobson and Lasenby we show that this macroscopic scalar field is not related to the mass generation that is required in a conformal theory. Rather, not just in conformal gravity, but also in standard Einstein gravity, the presence of such a long range scalar field  would lead to test particles whose masses would be of the same order as the masses of the galaxies around which they orbit.  Since particle masses are not at all of this form,  such macroscopic fields cannot be responsible for mass generation;  and the existence of any such mass-generating scalar fields can be excluded, consistent with there actually being no known massless scalar particles in nature. Instead, mass generation has to be due to c-number vacuum expectation values of q-number fields. Such expectation values are microscopic not macroscopic and only vary within particle interiors, giving particles an extended, baglike structure, as needed for localization in a conformal theory. And being purely internal they have no effect on galactic orbits, to thus leave the good conformal gravity fitting to galactic rotation curves intact.

\end{abstract}

\maketitle

%\tableofcontents
\section{Introduction}
\label{S1}

\subsection{Implications of mass generation for gravity}
\label{S1A}

Even though massive sources play a central role in theories of gravity, as long as the gravity theory is both classical and macroscopic, these masses can be treated as being purely mechanical or kinematical in nature. However, the actual generation of mass itself is intrinsically microscopic, being due to a quantum-field-theoretic, spontaneously broken Hilbert space vacuum in which a quantum field such as $\phi$ or a quantum fermion bilinear such as $\bar{\psi}\psi$ acquires a non-vanishing vacuum expectation value.  Nonetheless, it is generally thought that macroscopic gravity is not sensitive to such microscopic effects, with gravity only needing to respond to the presence of a massive source irrespective of how that mass may or may not have arisen. And yet once one has  an expectation value such as $S(x)=\langle \Omega |\phi|\Omega\rangle$ or $S(x)=\langle \Omega|\bar{\psi}\psi|\Omega \rangle$, one can ask whether this c-number $S(x)$ order parameter might itself extend to large distances and have macroscopic consequences. And if the vacuum is translation invariant the quantity $S(x)$ would be a  constant and thus would indeed extend throughout all spacetime, and not only would it do so, it would then provide a potentially substantial contribution to the cosmological constant. Moreover, if evaluated in a coherent state $|C\rangle$, a quantity such as $S(x)=\langle C|\phi|C\rangle$ or $S(x)=\langle C|\bar{\psi}\psi|C \rangle$ would then be spacetime dependent and could thus have a long range spacetime dependence. With such a spacetime-dependent  $S(x)$ carrying both energy and momentum it could contribute to the classical gravitational equations of motion in a possibly significant way.  In addition, if Yukawa coupled to a fermion via a Lagrangian density term of the form $-hS(x)\bar{\psi}(x)\psi(x)$, such an $S(x)$ could modify the motion of a fermionic particle over large distances scales, so that its trajectory (or that of a star composed of fermions) in a gravitational field would then not  be geodesic. 

To see the nature of the issue in detail consider a minimally coupled scalar field with wave equation $\nabla_{\mu}\nabla^{\mu}S=0$ propagating in a static, spherically symmetric geometry with line element $ds^2=B(r)dt^2-A(r)dr^2-r^2d\theta^2-r^2 \sin^2\theta d\phi^2$ in an arbitrary metric theory of gravity. In such a geometry the wave equation would take the form $(ABr^2)^{-1/2}\partial_r[r^2(B/A)^{1/2}\partial_rS]=0$, and the scalar  field would acquire a radial dependence of the form  $S(r)=c\int dr A^{1/2}/r^2B^{1/2}$ where $c$ is an integration constant. If the metric coefficients $B(r)$ and $A(r)$ are fixed by some static source such as a galaxy with $N^*$ stars so that $A(r)$ and $B(r)$ are of order $N^*$, $S(r)$ would accordingly acquire a term of order $N^*$. But that would lead to a Yukawa coupling to a single star orbiting the galaxy that would also be of order $N^*$. Such a coupling could not be ascribed to the mass generation mechanism for a single star as it would be $N^*$ times too large, and not only that, it would give an effect that would vary from one galaxy to the next as $N^*$ itself would then vary. 

Such a situation could for instance have been encountered in the scalar-tensor Brans-Dicke theory \cite{Brans1961},  in which a macroscopic  scalar field is coupled to Einstein gravity. However, Brans and Dicke only allowed the scalar field to couple to the gravitational sector and not to the matter sector. And as emphasized in \cite{Weinberg1972}, any such coupling to the matter sector could impair the successes of the equivalence principle. The Brans-Dicke theory thus finesses the issue by not associating the macroscopic scalar field with mass generation, i.e., the Brans-Dicke scalar field does not Yukawa couple to propagating fermions. In the Brans-Dicke theory the lack of any such Yukawa coupling is essentially imposed by fiat. However, the  typical Lagrangian density for a scalar field, viz.  $-[\nabla_{\mu}S\nabla^{\mu}S/2-m^2S^2/2+\lambda S^4]$,  has a discrete symmetry under $S \rightarrow -S$, and thus imposing this symmetry could actually forbid the presence of any  $-hS\bar{\psi}\psi$ term as it would be odd under the discrete symmetry if $\bar{\psi}\psi$ is even. Alternatively, the scalar field could be associated with some very high mass  scale such as a cosmological scale or  a grandunified scale,  scales that are not associated with low scale fermion masses.

In not Yukawa coupling the scalar field to the matter sector the Brans-Dicke theory avoids another problem. Specifically, while the c-number $S(x)$ might describe the vacuum or a coherent state, quantizing fluctuations around either such configuration would give rise to propagating quantum scalar fields, which if Yukawa coupled to fermions would mediate interactions between fermions, in direct analog to photon or graviton exchange. These interactions would then be long range if the quantized scalar field is massless. There appears to be no sign of any such long range interaction, and concomitantly, there actually appears to be no known scalar (or pseudoscalar) particle that is in fact massless. Thus for the $SU(2)\times U(1)$ electroweak theory for instance, while the vacuum expectation value of the Higgs field gives rise to fermion masses, the associated fluctuating scalar Higgs field that is excited out of the Higgs vacuum is actually massive, to thus lead to acceptably short range rather than long range interactions. In  addition, the electroweak Higgs field is accompanied by a set of massless Goldstone bosons, and initially the exchange of these particles could  give rise to long range interactions. However, via the Englert-Brout-Higgs mechanism these Goldstone bosons are incorporated into a matching set of gauge bosons that then become massive, with no massless Goldstone bosons being left in the spectrum. The electroweak theory thus nicely protects itself from  the presence of any long range interaction that could be mediated by spin zero scalar or pseudoscalar particles, with this all being done at the microscopic, not macroscopic, level.

While these remarks would apply to any metric theory of gravity, for the conformal gravity theory of interest to us in this paper there are additional considerations. Specifically, in a theory that has a conformal symmetry there are no mass scales at the level of the Lagrangian. Consequently, all mass must be generated in the vacuum, and thus both the stars in a galaxy and a test particle that is orbiting around the galaxy must all acquire their masses by one and the same mechanism. Thus unlike in the Brans-Dicke case, one cannot decouple the mass generation mechanism for the stars in a gravitational source (the stars that fix the above $A(r)$ and $B(r)$ coefficients) from the mass generation mechanism for the test star that orbits around the galaxy. Thus if the scalar field does couple to the stars in the source, it must couple to the test  particle too, and one is not free to ignore this coupling. Moreover, if one were to introduce a discrete $S\rightarrow -S$ symmetry, while it might then prevent the scalar field from coupling to the test particle, it would equally prevent the scalar field from  coupling to and giving mass to the particles in the source as well. 

An  additional, and quite unusual aspect of the conformal gravity theory is that even if there is a radially dependent macroscopic scalar field, according to \cite{Brihaye2009} it can actually hide itself by not contributing to the energy-momentum tensor in the region exterior to a static, spherically symmetric source. Its presence thus does not modify the exterior region solution for the metric that would have been obtained had it not been present at all. However, such a scalar field is not completely without consequence since, as noted in \cite{Horne2016} and \cite{Hobson2021}, its Yukawa coupling to a test particle (as realized via a test particle action of the form $I_T=-h\int S(x)ds$) would lead to a  trajectory for the test particle that is not geodesic.  Such non-geodesic behavior would represent a challenge to the dark-matter-free conformal gravity fits to 138 galactic rotation curves given in \cite{Mannheim2011,Mannheim2012,O'Brien2012}, fits that were based on the use of geodesic orbits. Thus in contrast to the Brans-Dicke study where the scalar field couples to the geometry but not to the test particle, in the conformal case the scalar field couples to the test particle but does not affect the geometry in the exterior region.

It is the purpose of this paper to revisit the analyses of \cite{Brihaye2009},  \cite{Horne2016} and \cite{Hobson2021}, and show that they actually reinforce the point we raised above, namely  that one cannot associate a radially varying, macroscopic scalar field with mass generation in the first place. Now initially this might not appear to be the case since the very fact that the scalar field causes the exterior region $T_{\mu\nu}$ to vanish means that the strength of the scalar field is not fixed by the number $N^*$ of stars in the source. However, it turns out that we need to reexamine the study made in \cite{Brihaye2009}. Specifically, for a general static, spherically symmetric energy-momentum tensor there are only three independent components: $T_{00}$, $T_{rr}$ and $T_{\theta\theta}$. However, for the traceless  $T_{\mu\nu}$ required of a conformal theory there are only two, and  as will become apparent below, the specific structure of the conformal gravitational equations actually make $T^{0}_{\phantom{0}0}-T^{r}_{\phantom{r}r}$ and $T^{r}_{\phantom{r}r}$ the optimal choice. And as noted in \cite{Brihaye2009}, the vanishing of $T^{0}_{\phantom{0}0}-T^{r}_{\phantom{r}r}$ in the region exterior to the matter source would force a classical, macroscopic scalar field to have the form $S(r)=1/(b+ar)$; with the very vanishing of $T^{0}_{\phantom{0}0}-T^{r}_{\phantom{r}r}$ allowing the parameters $a$ and $b$ to not be fixed by any matter fields that are located in the interior region. 

Now in their study Brihaye and Verbin and then subsequently Horne and then Hobson and Lasenby took  $T^{r}_{\phantom{r}r}$ to vanish as well. However, we shall show below that this form for  $S(r)$ does not in and of itself actually cause $T^{r}_{\phantom{r}r}$ to vanish. That this could in principle be the case is due to the fact that for the general line element $ds^2=B(r)dt^2-A(r)dr^2-r^2d\theta^2-r^2\sin^2\theta d\phi^2$  the covariant conservation of a traceless $T_{\mu\nu}$ leads to 
\begin{eqnarray}
\left(\frac{B^{\prime}}{2B}-\frac{1}{r}\right)(T^0_{{\phantom 0} 0}-T^r_{{\phantom r} r})-\left(\frac{d}{dr}+\frac{4}{r}\right)T^r_{{\phantom r} r}=0.
\label{1.1y}
\end{eqnarray}                                 
Consequently, at best the vanishing of $T^{0}_{\phantom{0}0}-T^{r}_{\phantom{r}r}$ could only constrain $T^{r}_{\phantom{r}r}$ to be of the form $A/r^4$ where $A$ is a constant. In our study below we will use the full set of equations of motion of the theory to show that $T^{r}_{\phantom{r}r}$ does not in fact vanish if $b=0$, but does vanish if $b\neq 0$. With $b\neq 0$ we now can set all the components of $T_{\mu\nu}$ to zero  and thus establish the result that had been assumed in \cite{Brihaye2009, Horne2016,Hobson2021} where the coefficient $b$ had been taken to be non-zero. In order to actually get $T^{r}_{\phantom{r}r}$ to vanish we have to use the equations of motion, and this now does lead us to a relation between $a$ and $b$ and the metric coefficients produced by the interior region matter source, to thus force $a$ and $b$ to be of order $N^*$ in the galactic case. (As noted in \cite{Horne2016}, one can obtain a relation between the $a$ and $b$ parameters and the metric coefficients through the scalar field equation of motion, but as we show below, this does not force $T^{r}_{\phantom{r}r}$ to vanish.)

While mass generation is microscopic, nonetheless it is of interest to examine what would happen in the conformal theory if mass generation were to lead to a scalar field $S(x)$ that did vary macroscopically and couple to a test particle with an action of the form $I_T=-h\int S(x)ds$. In this paper we shall examine this option in detail in the explicitly solvable model introduced in \cite{Brihaye2009}, and explicitly demonstrate that such a test particle action would give rise to masses for test stars that were not of order one solar mass but of order the mass of the galaxy around which they  orbit. Since it is thus not valid to ascribe mass generation to such a macroscopic field, the concerns raised in \cite{Horne2016} and \cite{Hobson2021} (studies that used this $I_T$ test particle action) are thus invalidated.

If mass generation is not to be associated with an  $S(x)$ that varies macroscopically then it must be associated with an $S(x)$ that only varies on microscopic scales while being constant on large scales. Structures in which the order parameter $S(x)$ only varies within the particles that it gives mass to are well established in flat spacetime dynamical theories where they are known as extended models or bag models of hadrons, i.e., models of hadrons with non-zero radii.  In particular, they can arise in flat spacetime theories that have no intrinsic dimensional parameters. And it was this particular aspect of mass generation that first led \cite{Mannheim1994,Mannheim1994a}  the present author to generalize these flat space concepts to curved space in a gravity theory, conformal gravity,  in which there again would be no intrinsic dimensional parameters.  Moreover, such extended structures are actually necessary in the conformal theory since the mechanism that generates mass dynamically generates extended structures at the same time, to thereby enable particles to localize into the stars and galaxies that are of relevance to astrophysics and cosmology. Moreover, in these dynamical models coherent states such as $|C\rangle$ are in and of themselves not actually stable. (They would be stable if they had a non-trivial topology, though that is not the case here.) The coherent states only become stable in the presence of a positive energy fermion excited out of the filled negative energy Dirac sea. And in this stabilization mechanism the fermion becomes localized and the scalar field only varies from a constant value within the localization region. Thus the very nature of dynamical mass generation is such that there can be no macroscopic order parameters that vary radially on large distance scales. This is completely consistent with our showing in this paper that one could not ascribe mass generation to macroscopic scalar fields that do vary on large distance scales.

This paper is organized as follows. After introducing the conformal theory in the rest of this section, in Sec. \ref{S2} we present the exact solution  to the conformal theory in the static, spherically symmetric case in which there are no matter  fields in the exterior region. In this analysis we present the linear potentials that are key to the conformal gravity theory fitting to galactic rotation curves, noting that there are actually two linear potentials of relevance, one due to the stars inside a local inhomogeneous matter source such as a galaxy and the other being due to the homogeneous global background cosmology. In Sec. \ref{S3} we follow \cite{Brihaye2009} and augment the theory with a scalar field that is to be present in this exterior region, and investigate its implications for the behavior of  the scalar field $T_{\mu\nu}$. This analysis enables us to establish that the linear potential remains in the solution and that the $a$ and $b$ parameters in the $S(r)=1/(b+ar)$ solution are of galactic and even cosmological scale. In Sec. \ref{S4} we discuss the implications of coupling the same scalar field to a test particle. As noted in \cite{Horne2016} the linear potential contribution is removed from the test particle orbit in the $S=1/(b+ar)$ case, while it is not present at all in the $S=1/b$ case, with the effect of the linear potential thus being lost in both of the cases. However, in either case we establish that at the same time this would lead to a test particle that would acquire an unacceptable galactic or cosmological sized mass.  In addition, we identify some inconsistencies in the calculation given in \cite{Horne2016}. Moreover, while, as noted in \cite{Horne2016,Hobson2021}, one can actually remove the key linear potential by a conformal transformation when there is just a single source and radial symmetry, in Sec. \ref{S5} we show that this is not the case for a multiparticle source such as a spiral galaxy with axial not radial symmetry. Finally, in Sec. \ref{S6} we show that through use of the microscopic scalar field $\langle C|\bar{\psi}\psi|C \rangle$ that is appropriate to dynamical symmetry breaking test particles then move on standard geodesics. That this is the case is because in the self-consistent vacuum the conformal symmetry is broken in both the vacuum and the action, with the fermion that is excited out of the vacuum not being constrained to propagate according to a conformal invariant  equation of motion, and with the linear potential term then not being removed even though the scalar field is macroscopically constant. In an appendix we follow \cite{Mannheim2006} and discuss the cosmological origin of the global linear potential, with there being a conformal and coordinate equivalence  between the static, spherically symmetric geometries of relevance to galactic rotation curves and the comoving Robertson-Walker geometry of relevance to cosmology. 

\subsection{The gravitational sector of conformal gravity}
\label{S1A1}

Conformal gravity has been advanced as a candidate alternate theory of gravity (see e.g. the reviews \cite{Mannheim2006,Mannheim2017} and references therein). The theory is a pure metric theory of gravity that possesses all of the general coordinate invariance and equivalence principle structure of standard gravity while augmenting it with an additional symmetry, local conformal invariance, in which  the action is left invariant under local conformal transformations on the metric of the form $g_{\mu\nu}(x)\rightarrow e^{2\alpha(x)}g_{\mu\nu}(x)$ with arbitrary local phase $\alpha(x)$. Under such a symmetry a gravitational sector action that is to be a polynomial function of the Riemann tensor is uniquely prescribed, and with use of the Gauss-Bonnet theorem is given by (see e.g. \cite{Mannheim2006}) 
\begin{eqnarray}
I_{\rm W}=-\alpha_g\int d^4x\, (-g)^{1/2}C_{\lambda\mu\nu\kappa}
C^{\lambda\mu\nu\kappa}
\equiv -2\alpha_g\int d^4x\, (-g)^{1/2}\left[R_{\mu\kappa}R^{\mu\kappa}-\frac{1}{3} (R^{\alpha}_{\phantom{\alpha}\alpha})^2\right].
\label{1.2y}
\end{eqnarray}
Here $\alpha_g$ is a dimensionless  gravitational coupling constant, and
\begin{eqnarray}
C_{\lambda\mu\nu\kappa}= R_{\lambda\mu\nu\kappa}
-\frac{1}{2}\left(g_{\lambda\nu}R_{\mu\kappa}-
g_{\lambda\kappa}R_{\mu\nu}-
g_{\mu\nu}R_{\lambda\kappa}+
g_{\mu\kappa}R_{\lambda\nu}\right)
+\frac{1}{6}R^{\alpha}_{\phantom{\alpha}\alpha}\left(
g_{\lambda\nu}g_{\mu\kappa}-
g_{\lambda\kappa}g_{\mu\nu}\right)
\label{1.3y}
\end{eqnarray}
is the conformal Weyl tensor. (Here and throughout unless indicated to the contrary we follow the notation and conventions of \cite{Weinberg1972}.) 

With the Weyl action $I_{\rm W}$ given in  (\ref{1.2y}) being a fourth-order derivative function of the metric, functional variation with respect to the metric $g_{\mu\nu}(x)$ generates fourth-order derivative gravitational equations of motion of the form \cite{Mannheim2006} 
\begin{eqnarray}
-\frac{2}{(-g)^{1/2}}\frac{\delta I_{\rm W}}{\delta g_{\mu\nu}}=4\alpha_g W^{\mu\nu}=4\alpha_g\left[2\nabla_{\kappa}\nabla_{\lambda}C^{\mu\lambda\nu\kappa}-
R_{\kappa\lambda}C^{\mu\lambda\nu\kappa}\right]=4\alpha_g\left[W^{\mu
\nu}_{(2)}-\frac{1}{3}W^{\mu\nu}_{(1)}\right]=T^{\mu\nu},
\label{1.4y}
\end{eqnarray}
where the functions $W^{\mu \nu}_{(1)}$ and $W^{\mu \nu}_{(2)}$ (respectively associated with the $(R^{\alpha}_{\phantom{\alpha}\alpha})^2$ and $R_{\mu\kappa}R^{\mu\kappa}$ terms in (\ref{1.2y})) are given by
\begin{eqnarray}
W^{\mu \nu}_{(1)}&=&
2g^{\mu\nu}\nabla_{\beta}\nabla^{\beta}R^{\alpha}_{\phantom{\alpha}\alpha}                                             
-2\nabla^{\nu}\nabla^{\mu}R^{\alpha}_{\phantom{\alpha}\alpha}                          
-2 R^{\alpha}_{\phantom{\alpha}\alpha}R^{\mu\nu}                              
+\frac{1}{2}g^{\mu\nu}(R^{\alpha}_{\phantom{\alpha}\alpha})^2,
\nonumber\\
W^{\mu \nu}_{(2)}&=&
\frac{1}{2}g^{\mu\nu}\nabla_{\beta}\nabla^{\beta}R^{\alpha}_{\phantom{\alpha}\alpha}
+\nabla_{\beta}\nabla^{\beta}R^{\mu\nu}                    
 -\nabla_{\beta}\nabla^{\nu}R^{\mu\beta}                       
-\nabla_{\beta}\nabla^{\mu}R^{\nu \beta}                          
 - 2R^{\mu\beta}R^{\nu}_{\phantom{\nu}\beta}                                    
+\frac{1}{2}g^{\mu\nu}R_{\alpha\beta}R^{\alpha\beta},
\label{1.5y}
\end{eqnarray}                                 
and where $T^{\mu\nu}$ is the conformal invariant and thus traceless energy-momentum tensor associated with a conformal matter source. With the action  being both general coordinate invariant and conformal invariant $W^{\mu\nu}$ is automatically covariantly conserved and covariantly traceless (i.e., it obeys $\nabla_{\nu}W^{\mu\nu}=0$, $g_{\mu\nu}W^{\mu\nu}=0$) without the need to impose any gravitational equation of motion. And with $I_{\rm W}$ being conformal invariant $W_{\mu\nu}$ transforms as $W_{\mu\nu}\rightarrow e^{-2\alpha(x)}W_{\mu\nu}$ under a local conformal transformation. In  addition the Weyl tensor vanishes in geometries that are conformal to flat, so that in the homogeneous and isotropic conformal to flat background Robertson-Walker and de Sitter geometries of interest to cosmology the background $W^{\mu\nu}$ is zero. Despite this,  fluctuations (i.e., cosmological inhomogeneities such as the galaxies and stars of interest to us in this paper) around that background are not conformal to flat, with the fluctuating $\delta W^{\mu\nu}$ then not being zero. 

\subsection{The matter sector}
\label{S1B}

Since the light cone is left invariant under a local conformal transformation, in a conformal invariant theory there are no mass scales at the level of the Lagrangian, with  all particles being massless.  The generation of fermion mass scales in a conformal invariant theory has to be done in the vacuum, and is characterized by an order parameter $S(x)$ (a vacuum expectation value). For the fermion and scalar field sector we take the matter action to be of the form 
\begin{eqnarray}
I_M=-\int d^4x(-g)^{1/2}\left[\frac{1}{2}\nabla_{\mu}S\nabla^{\mu}S-\frac{1}{12}S^2R^\mu_{\phantom{\mu}\mu}
+\lambda S^4+
i\bar{\psi}\gamma^{c}V^{\mu}_c(x)[\partial_\mu+\Gamma_\mu(x)]\psi-hS\bar{\psi}\psi\right],
\label{1.6y}
\end{eqnarray}                                 
where $h$ and $\lambda$ are dimensionless coupling constants.  Here $V^{\mu}_c(x)$ is a vierbein and $\Gamma_{\mu}=-(1/8)[\gamma_a,\gamma_b](V^b_{\nu}\partial_{\mu}V^{a\nu}+V^b_{\lambda}\Gamma^{\lambda}_{\phantom{\lambda}\nu\mu}V^{a\nu})$ is the spin connection, with the $\gamma_a$ being a set of fixed axis Dirac gamma matrices and $\Gamma^{\lambda}_{\phantom{\lambda}\nu\mu}$ being the Levi-Civita connection $(1/2)g^{\lambda\sigma}[\partial_{\mu}g_{\sigma\nu} +\partial_{\nu}g_{\sigma\mu} -\partial_{\sigma}g_{\mu\nu} ]$. As such, this action is locally conformal invariant under $g_{\mu\nu}(x)\rightarrow e^{2\alpha(x)}g_{\mu\nu}(x)$, $V^a_{\mu}(x)\rightarrow e^{\alpha(x)}V^a_{\mu}(x)$, $\psi(x)\rightarrow e^{-3\alpha(x)/2}\psi(x)$, $S(x)\rightarrow e^{-\alpha(x)}S(x)$.  Variation of
this action with respect to  $\psi(x)$ and $S(x)$ yields the matter field equations of motion
\begin{eqnarray}
i\gamma^{c}V^{\mu}_c(x)[\partial_\mu+\Gamma_\mu(x)]\psi - h S \psi = 0,
\label{1.7y}
\end{eqnarray}                                 
\begin{eqnarray}
\nabla_{\mu}\nabla^{\mu}S+\frac{1}{6}SR^\mu_{\phantom{\mu}\mu}
-4\lambda S^3 +h\bar{\psi}\psi=0,
\label{1.8y}
\end{eqnarray}                                 
while variation with respect to the metric yields an energy-momentum tensor of the form (first without use of any matter field equation of motion and then with)
\begin{eqnarray}
T^{\mu \nu}&=&i\bar{\psi}\gamma^{c}V_{\mu c}(x)[\partial_{\nu}+\Gamma_{\nu}(x)]\psi +\frac{2}{3}\nabla^{\mu}S \nabla^{\nu} S
-\frac{1}{6}g^{\mu\nu}\nabla_{\alpha}S\nabla^{\alpha}S
-\frac{1}{3}S\nabla^{\mu}\nabla^{\nu}S
\nonumber \\             
&+&\frac{1}{3}g^{\mu\nu}S\nabla_{\alpha}\nabla^{\alpha}S                           
-\frac{1}{6}S^2\left(R^{\mu\nu}
-\frac{1}{2}g^{\mu\nu}R^\alpha_{\phantom{\alpha}\alpha}\right)
\nonumber\\
&-&g^{\mu\nu}[\lambda S^4+i\bar{\psi}\gamma^{c}V^{\mu}_c(x)[\partial_\mu+\Gamma_\mu(x)]\psi - h S\bar{\psi} \psi]
\nonumber\\
&=&i\bar{\psi}\gamma^{c}V_{\mu c}(x)[\partial_{\nu}+\Gamma_{\nu}(x)]\psi +\frac{2}{3}\nabla^{\mu}S \nabla^{\nu} S
-\frac{1}{6}g^{\mu\nu}\nabla_{\alpha}S\nabla^{\alpha}S
-\frac{1}{3}S\nabla^{\mu}\nabla^{\nu}S
\nonumber \\             
&+&\frac{1}{3}g^{\mu\nu}S\nabla_{\alpha}\nabla^{\alpha}S                           
-\frac{1}{6}S^2\left(R^{\mu\nu}
-\frac{1}{2}g^{\mu\nu}R^\alpha_{\phantom{\alpha}\alpha}\right)-g^{\mu\nu}\lambda S^4.
\label{1.9y}
\end{eqnarray}                                 
Use of the matter field equations of motion then confirms that the matter sector energy-momentum tensor is indeed traceless. Moreover, from the conformal invariance of $I_{\rm M}$ it follows that $T_{\mu\nu}$ transforms as $T_{\mu\nu}\rightarrow e^{-2\alpha(x)}T_{\mu\nu}$ under a local conformal transformation. With $W_{\mu\nu}$ transforming the same way, the $4\alpha_gW_{\mu\nu}=T_{\mu\nu}$ equation of motion is conformal invariant.
We now proceed to look for static, spherically symmetric solutions to these equations. As such, these selfsame equations could be considered to be microscopic or macroscopic, and so we first explore them as macroscopic equations, and defer a discussion of them as microscopic ones to Sec. \ref{S6}.

\section{The Equations of Motion in the Static, Spherically Symmetric Case}
\label{S2}

For a static, spherically symmetric source the Weyl tensor, and thus $W^{\mu\nu}=2\nabla_{\kappa}\nabla_{\lambda}C^{\mu\lambda\nu\kappa}-R_{\kappa\lambda}C^{\mu\lambda\nu\kappa}$,  do not vanish, and thus in trying to solve the  equation $W^{\mu\nu}=T^{\mu\nu}/4\alpha_g$, we have to deal with a fourth-order differential equation. As noted in \cite{Mannheim1989}, it turns out that in the static, spherically symmetric case it is possible to greatly simplify the problem through use of the underlying conformal symmetry that the theory possesses. Specifically,  under the general coordinate transformation
\begin{eqnarray}
\rho = p(r), \qquad B(r) = \frac{r^2 b(r)}{p^2(r)},\qquad                             
 A(r) =\frac{r^2 a(r) p^{\prime 2}(r)}{p^2 (r)},
\label{2.1x}
\end{eqnarray}
with an initially arbitrary function $p(r)$, the general static, spherically symmetric line element
\begin{eqnarray}
ds^2 = b(\rho)dt^2 - a(\rho)d\rho ^2 - \rho^2d\theta^2-\rho^2\sin^2\theta d\phi^2
\label{2.2x}
\end{eqnarray}
is brought to the form
\begin{eqnarray}
ds^2 = \frac{p^2(r)}{ r^2}\left[B(r) dt^2 - A(r) dr^2 - r^2d\theta^2-r^2\sin^2\theta d\phi^2\right].  
\label{2.3x}
\end{eqnarray}
On now choosing $p(r)$ according to 
\begin{eqnarray}
-\frac{1}{p(r)} = \int^r\frac{dr}{r^{2}[a(r)b(r)]^{1/2}},
\label{2.4x}
\end{eqnarray}
the function $A(r)$ would then be set equal to $1/B(r)$, with the 
line element then being brought to the 
form                                               
\begin{eqnarray}
ds^2= \frac{p^2(r)}{ r^2}\left[B(r) dt^2- \frac{dr^2}{B(r)} - r^2d\theta^2-r^2\sin^2\theta d\phi^2\right].
\label{2.5x}
\end{eqnarray}
While (\ref{2.5x}) is coordinate equivalent to (\ref{2.2x}), the utility of (\ref{2.5x}) lies in the
fact that one of the two functions present in it  appears purely as  an overall
multiplier. Consequently, since both $W_{\mu\nu}$ and $T_{\mu\nu}$ transform the same way under a local conformal transformation  the function $p(r)/r$ can be removed from the
theory altogether (i.e., gauged away) via a local conformal transformation, with
the full kinematic content of the conformal theory then being contained in
the line element
\begin{eqnarray}
ds^2 = B(r)dt^2 -\frac{dr^2}{B(r)} -r^2d\theta^2-r^2\sin^2\theta d\phi^2.
\label{2.6x}
\end{eqnarray}

Given (\ref{2.6x}) we can determine the components of $W_{\mu\nu}$ in a closed form. With the trace condition $-BW^{00}+(1/B)W^{rr}+2r^2W^{\theta\theta}=0$ holding, we only need to express $W^{00}$ and $W^{rr}$ in terms of $B(r)$. In  \cite{Mannheim1989} it was shown that $W^r_{{\phantom r} r}$ is given without any approximation at all as
\begin{eqnarray}
W^r_{{\phantom r} r}=\frac{1}{3r^4}(1+y^3y^{\prime\prime}),
\label{2.7x}
\end{eqnarray}
where $y^2=r^2B^{\prime}-2rB$. Then in \cite{Mannheim1994} it was shown that
\begin{eqnarray}                                                                               
\frac{3}{B}\left(W^0_{{\phantom 0} 0}-W^r_{{\phantom r} r}\right)=B^{\prime\prime\prime\prime}+\frac{4}{r}B^{\prime\prime\prime}.
\label{2.8x}
\end{eqnarray}      
(The right-hand side of (\ref{2.8x}) is the radial component of $\vec{\nabla}^4B$.) Both (\ref{2.7x}) and (\ref{2.8x}) are remarkably compact.

In the presence of a source the relevant components of (\ref{1.4y}) take the form 
\begin{eqnarray}                                                                               
B^{\prime\prime\prime\prime}+\frac{4}{r}B^{\prime\prime\prime}&=&\frac{3}{4\alpha_g B}\left(T^0_{{\phantom 0} 0}
-T^r_{{\phantom r} r}\right) =f(r),
\label{2.9x}
\end{eqnarray}      
\begin{eqnarray}                                                                               
\frac{1}{3r^4}(1+y^3y^{\prime\prime})=\frac{T^r_{{\phantom r} r}}{4\alpha_g},
\label{2.10x}
\end{eqnarray}      
with (\ref{2.9x}) serving to define $f(r)$.

The solution to (\ref{2.9x}) can be determined in closed form and is given by  \cite{Mannheim1994} 
\begin{eqnarray}
B(r)&=&B_0(r)-\frac{r}{2}\int_0^r
dr^{\prime}r^{\prime 2}f(r^{\prime})
-\frac{1}{6r}\int_0^r
dr^{\prime}r^{\prime 4}f(r^{\prime})
-\frac{1}{2}\int_r^{\infty}
dr^{\prime}r^{\prime 3}f(r^{\prime})
-\frac{r^2}{6}\int_r^{\infty}
dr^{\prime}r^{\prime }f(r^{\prime}).
\label{2.11x}
\end{eqnarray}                                 
In (\ref{2.11x}) we have included an allowable  $B_0(r)$ contribution that identically satisfies $B_0^{\prime\prime\prime\prime}+(4/r)B_0^{\prime\prime\prime}=0$ at all points $r$, i.e., including those where $f(r)$ is non-zero. 

To constrain the $B_0(r)$ term we note that the $f(r)$-dependent term is due to sources that are associated with an expressly non-vanishing $W_{\mu\nu}$ (i.e., inhomogeneities in the cosmological background), while the $B_0(r)$ term is associated with sources for which $W_{\mu\nu}$ expressly vanishes everywhere (viz. the cosmological background itself). There are two ways in which the fourth-order derivative function $W_{\mu\nu}$ could vanish identically everywhere: the Weyl tensor itself could vanish or the Weyl tensor could  obey $2\nabla_{\kappa}\nabla_{\lambda}C^{\mu\lambda\nu\kappa}-R_{\kappa\lambda}C^{\mu\lambda\nu\kappa}=0$. With the non-vanishing components of of the Weyl tensor being given by $ B^{\prime\prime}-2B^{\prime}/r+2(B-1)/r^2$, for vanishing $C^{\mu\lambda\nu\kappa}$ the solution is given as $B_0(r)=1+\gamma_0r-K_0r^2$ where, as discussed in the appendix,  $\gamma_0$ and $K_0$ are universal constants that are associated with the cosmological background. For $2\nabla_{\kappa}\nabla_{\lambda}C^{\mu\lambda\nu\kappa}-R_{\kappa\lambda}C^{\mu\lambda\nu\kappa}=0$ the two other solutions required  for the fourth-order derivative $W_{\mu\nu}=0$ equation are given by $B_0(r)=w_0+u_0/r$. Since this solution has to hold everywhere a finiteness condition at $r=0$ excludes the $u_0/r$ term. Hence in the following  we set $B_0(r)=1+w_0+\gamma_0r-K_0r^2$. On defining $B^*(r)$ as the $f(r)$-dependent component of $B(r)$ that is given in (\ref{2.11x}), we can express the full $B(r)$ as $B(r)=B_0(r)+B^*(r)$.

From (\ref{2.11x}) it follows that  (\ref{2.10x}) takes the form (see e.g. the analogous discussion in \cite{Horne2016})                      
\begin{eqnarray}                                 
\frac{1}{12r^4}\left[4+
\int_0^rdr^{\prime}r^{\prime 2}f(r^{\prime})\int_0^rdr^{\prime}r^{\prime 4}f(r^{\prime}) 
-\left(\int_r^{\infty}dr^{\prime}r^{\prime 3}f(r^{\prime})-2w_0-2\right)^2
-2\gamma_0\int_0^rdr^{\prime}r^{\prime 4}f(r^{\prime})\right]
=\frac{T^r_{{\phantom r} r}}{4\alpha_g}.
\label{2.12x}
\end{eqnarray}                                 
On evaluating the non-vanishing components of the Weyl tensor for the $B(r)$ given in (\ref{2.11x}) we obtain
\begin{eqnarray}
B^{\prime\prime}-\frac{2}{r}B^{\prime}+\frac{2}{r^2}(B-1)=-\frac{1}{r^3}\int_0^rdr^{\prime}r^{\prime 4}f(r^{\prime})-\frac{1}{r^2}\int_r^{\infty}
dr^{\prime}r^{\prime 3}f(r^{\prime})+\frac{2w_0}{r^2},
\label{2.13x}
\end{eqnarray}
to thus confirm that the non-vanishing of the matter source $f(r)$ leads to a non-vanishing Weyl tensor.

For the metric given in (\ref{2.6x}) the various components of the Ricci tensor take the form
\begin{align}
R_{rr}&=\frac{B^{\prime\prime}}{2B}+\frac{B^{\prime}}{rB},\quad R_{\theta\theta}=-1+rB^{\prime}+B,\quad
R_{\phi\phi}=\sin^2\theta R_{\theta\theta},
\nonumber\\
R_{tt}&=-\frac{B^{\prime\prime}B}{2}-\frac{B^{\prime}B}{r},\quad
R^{\alpha}_{\phantom{\alpha}\alpha}=B^{\prime\prime}+\frac{4B^{\prime}}{r}+\frac{2B}{r^2}-\frac{2}{r^2}.
\label{2.14x}
\end{align}
Thus in the solution given in (\ref{2.11x}) the Ricci scalar is given by
\begin{align}
R^{\alpha}_{\phantom{\alpha}\alpha}&=-\frac{3}{r}\int_0^r
dr^{\prime}r^{\prime 2}f(r^{\prime})-\frac{1}{r^2}\int_r^{\infty}
dr^{\prime}r^{\prime 3}f(r^{\prime})-2\int_r^{\infty}
dr^{\prime}r^{\prime }f(r^{\prime})-12K_0+\frac{6\gamma_0}{r}
+\frac{2w_0}{r^2}.
\label{2.15x}
\end{align}

Some simplification of the general $B(r)$ is possible depending on how the matter sources are distributed. Since $f(r)$ is produced by inhomogeneities in the cosmological background, $f(r)$ will get contributions from both nearby matter and distant matter. The nearby matter is associated with any given galaxy of interest and the distant matter will be composed of large structures such as clusters of galaxies. While we will need to take the spiral (i.e., non-spherical) nature of galaxies composed of $N^*$ stars into consideration below,  for the moment we shall take the nearby matter to consist of a static, spherically symmetric source of radius $R_0$ with its center situated at the origin of the coordinate system that we are using,  so that its contribution to $f(r)$ is only non-zero in an $r<R_0$ region. Similarly, we shall take the large scale structure contribution to begin at some cluster of galaxies scale $R_1$. There is thus a region $R_0<r<R_1$ in which $f(r)$ is zero. In this region $B(r)$ takes the form
\begin{eqnarray}
B(R_0<r<R_1)&=&1+w_0+\gamma_0r-K_0r^2
-\frac{1}{6r}\int_0^{R_0}dr^{\prime}r^{\prime 4}f(r^{\prime})
-\frac{r}{2}\int_0^{R_0}dr^{\prime}r^{\prime 2}f(r^{\prime})
\nonumber\\
&-&\frac{1}{2}\int_{R_1}^{\infty}dr^{\prime}r^{\prime 3}f(r^{\prime})
-\frac{r^2}{6}\int_{R_1}^{\infty}dr^{\prime}r^{\prime }f(r^{\prime}).
\label{2.16x}
\end{eqnarray}                                 
On defining 
\begin{eqnarray}
-2\beta^*&=& -\frac{1}{6}\int_0^{R_0}
dr^{\prime}r^{\prime 4}f(r^{\prime}),\qquad
\gamma^*=-\frac{1}{2}\int_0^{R_0}
dr^{\prime}r^{\prime 2}f(r^{\prime}),\qquad
w^*=-\frac{1}{2}\int_{R_1}^{\infty}dr^{\prime}r^{\prime 3}f(r^{\prime}),\quad
\kappa^*=\frac{1}{6}\int_{R_1}^{\infty}dr^{\prime}r^{\prime }f(r^{\prime}),
\nonumber\\
w&=&1+w_0+w^*,\quad u=-2\beta^*, \quad v=\gamma_0+\gamma^*, \quad K=K_0+\kappa^*,
\label{2.17x}
\end{eqnarray}                                 
we can write (\ref{2.16x}) as 
\begin{eqnarray}
B(R_0<r<R_1)=w-Kr^2+\frac{u}{r}+v r,
\label{2.18x}
\end{eqnarray}
to thus possess Newtonian, linear, and quadratic potentials. The $\beta^*$ and $\gamma^*$ terms (per star) are fixed by the behavior of $f(r)$ in the $r<R_0$ region (and would be replaced by $N^*\beta^*$ and $N^*\gamma^*$ for a galaxy with $N^*$ stars), while the $w^*$ and $\kappa^*$ terms  are fixed by the behavior of $f(r)$ in the $r>R_1$ region. All  four of the $\beta^*$, $\gamma^*$, $w^*$ and $\kappa^*$  terms would be zero unless $f(r)$ is non-vanishing in the relevant regions, i.e., only if, as per (\ref{2.13x}),  the Weyl tensor is non-vanishing in those regions too. While the quadratic $-\kappa^* r^2$ term has the same form as a $-K_0r^2$ de Sitter term we should note that they are not equivalent as the $-\kappa^* r^2$ term only has the form that it does in the $r<R_1$ region, while the de Sitter form holds for all $r$. For the $-\kappa^* r^2$ term  we must use the full $-(r^2/6)\int_r^{\infty} dr^{\prime}r^{\prime }f(r^{\prime})$ term given in (\ref{2.11x}) in the $r>R_1$ region. The $-\kappa^* r^2$ term is associated with a non-vanishing Weyl tensor, while the $-K_0r^2$ term is associated with a Weyl tensor that vanishes.

As constructed, the existence of the solution given in  (\ref{2.18x}) with its dimensionful $u$,  $v$ and $K$ parameters is indicative of spontaneous symmetry breaking, since the very existence of solutions to a set of equations that have lower symmetry than the equations themselves is characteristic of spontaneous symmetry breaking. Since the underlying theory in the conformal gravity case is conformal invariant (and thus scaleless), the existence of a solution to the theory that then has a scale can only be brought about by spontaneous symmetry breaking in a Hilbert space with a spontaneously broken vacuum. This would then be realized by the  $S(x)$ field provided it is not an elementary field but is instead associated with a c-number expectation value of appropriate q-number fields.

In the exterior $R_0<r<R_1$ region inserting (\ref{2.18x}) into (\ref{2.10x}) yields
\begin{eqnarray}
\frac{1}{3r^4}(1+y^3y^{\prime\prime})=\frac{1}{3r^4}\left[1+3uv-w^2\right]=\frac{T^r_{{\phantom r} r}(R_0<r<R_1)}{4\alpha_g}.
\label{2.19x}
\end{eqnarray}                                 
If we now take $T^r_{{\phantom r} r}$ to vanish in $R_0<r<R_1$ (i.e., for  the moment with no matter fields in that region), then from (\ref{2.19x}) we fix
\begin{eqnarray}
w^2=1+3uv,
\label{2.20x}
\end{eqnarray}                                 
and up to a renaming of the coefficients given in \cite{Mannheim1989} we recover the exterior solution given there, viz.
\begin{eqnarray}
B(R_0<r<R_1)=(1+3uv)^{1/2}+\frac{u}{r}+v r -Kr^2,
\label{2.21x}
\end{eqnarray}                                 
a solution that may also be found in \cite{Riegert1984b}. In terms of the parameters given in (\ref{2.17x}) the $w^2=1+3uv$ relation takes the form $(w_0+w^*)^2+2w_0+2w^*+6\gamma^*\beta^*+6\gamma_0\beta^*=0$, with this relation following from (\ref{2.12x}) in any region in which $f(r)$ vanishes provided $T^r_{\phantom{r}r}$ also vanishes in that region. It was for a metric of the form given in (\ref{2.21x}) that the conformal theory was successfully applied to galactic rotation curves in \cite{Mannheim2011,Mannheim2012,O'Brien2012}, with fixed fitted parameters $\beta^*=1.48\times 10^5~{\rm cm}$ per solar mass, $\gamma^*=5.42\times 10^{-41}~{\rm cm}^{-1}$ per solar mass, $\gamma_0=3.06\times 10^{-30}~{\rm cm}^{-1}$(viz. explicitly a cosmological scale), and $ \kappa^*=9.54\times 10^{-54}~{\rm cm}^{-2}$ (viz. explicitly a cluster of galaxies scale) for all the galactic rotation curves in the entire 138 galaxy sample, and with the particles orbiting in the galaxies being taken to be geodesic.

The derivation leading to (\ref{2.21x}) presupposed that there was no scalar field present in the exterior $R_0<r<R_1$  region. For (\ref{2.21x})  to continue to hold when there is scalar field in that region requires that  the behavior of $S(r)$ in the $R_0<r<R_1$ region be such that it does in fact cause both $T^0_{{\phantom 0} 0}-T^r_{{\phantom r} r}$ and $T^r_{{\phantom r} r}$ to vanish there.  We thus now explore the degree  to which  this is or is not the case.

\section{Solving the Equations of Motion with a Macroscopic Scalar Field}
\label{S3}

\subsection{The general case}
\label{S3A}

To obtain an exact solution to the equations of motion in the exterior $R_0<r<R_1$ region in the case when there are non-vanishing fields in that region, we restrict the source to just a macroscopic scalar field $S(x)$.  Following \cite{Brihaye2009} we look to see if we can have a non-trivial scalar field and yet nonetheless still get  $T_{\mu\nu}$ to vanish in the exterior  $R_0<r<R_1$ region.  Thus in the static, spherically symmetric case we take $S(x)$ to depend on the radial coordinate $r$. Then, given the metric in (\ref{2.6x}) and a scalar sector of the form that (\ref{1.8y}) and (\ref{1.9y}) would take if the fermion sector is excluded, in the exterior  $R_0<r<R_1$ region we obtain 
\begin{eqnarray}
T^0_{{\phantom 0} 0}-T^r_{{\phantom r} r}= \frac{B}{3}(SS^{\prime\prime}-2S^{\prime 2}),
\label{3.1x}
\end{eqnarray}                                 
\begin{eqnarray}
T^r_{{\phantom r} r}=\frac{B}{2}S^{\prime 2}+\frac{B^{\prime}}{6}SS^{\prime}+\frac{2B}{3r}SS^{\prime}+\frac{B^{\prime}}{6r}S^2+\frac{B}{6r^2}S^2 -\frac{1}{6r^2}S^2-\lambda S^4,
\label{3.2x}
\end{eqnarray}                                 
while the pure scalar sector of (\ref{1.8y}) becomes
\begin{eqnarray}
BS^{\prime\prime}+B^{\prime}S^{\prime}+\frac{2B}{r}S^{\prime}+\frac{S}{6}\left(B^{\prime\prime}+\frac{4B^{\prime}}{r}+\frac{2B}{r^2}-\frac{2}{r^2}\right)-4\lambda S^3=0.
\label{3.3x}
\end{eqnarray}                                 

To find an exterior solution it was suggested in \cite{Brihaye2009} to set $T^0_{{\phantom 0} 0}-T^r_{{\phantom r} r}=0$ \cite{footnoteC}.  This then requires that $SS^{\prime\prime}=2S^{\prime 2}$, with  integrals
\begin{eqnarray}
S^{\prime}=-aS^2,\qquad S(r)=\frac{1}{b+ar},
\label{3.4x}
\end{eqnarray}                                 
where $a$ and $b$ are integration constants.  If we insert (\ref{3.3x}) into (\ref{3.2x}) we obtain 
\begin{eqnarray}
T^r_{{\phantom r} r}=-\frac{B}{4}SS^{\prime\prime}+\frac{B}{2}S^{\prime 2}-\frac{B^{\prime}}{12}SS^{\prime}+\frac{B}{6r}SS^{\prime}-\frac{B^{\prime\prime}}{24}S^2+\frac{B}{12r^2}S^2 -\frac{1}{12r^2}S^2.
\label{3.5y}
\end{eqnarray}                                 
With (\ref{3.4x}) we can rewrite (\ref{3.5y}) as 
\begin{eqnarray}
T^r_{{\phantom r} r}=a\frac{B^{\prime}}{12}S^3-a\frac{B}{6r}S^3-\frac{B^{\prime\prime}}{24}S^2+\frac{B}{12r^2}S^2 -\frac{1}{12r^2}S^2.
\label{3.6y}
\end{eqnarray}                                 
Now the intent of \cite{Brihaye2009},  \cite{Horne2016} and \cite{Hobson2021} was to have the vanishing of $T^0_{{\phantom 0} 0}-T^r_{{\phantom r} r}$ cause $T^r_{{\phantom r} r}$ to vanish as well. As we see from (\ref{3.6y}), for an as yet unspecified form for $B(r)$ this is not the case (and not even if we were to set $a=0$). To see why $T^r_{{\phantom r} r}$ could not vanish even in principle absent further information of some sort,  we recall that we had noted above that because of its tracelessness and covariant conservation $T_{\mu\nu}$  and thus equally $W_{\mu\nu}$ obey
\begin{eqnarray}
\left(\frac{B^{\prime}}{2B}-\frac{1}{r}\right)(T^0_{{\phantom 0} 0}-T^r_{{\phantom r} r})-\left(\frac{d}{dr}+\frac{4}{r}\right)T^r_{{\phantom r} r}=0,\qquad
\left(\frac{B^{\prime}}{2B}-\frac{1}{r}\right)(W^0_{{\phantom 0} 0}-W^r_{{\phantom r} r})-\left(\frac{d}{dr}+\frac{4}{r}\right)W^r_{{\phantom r} r}=0.
\label{3.7y}
\end{eqnarray}                                 
Thus from (\ref{3.7y}) we see that even if we set $T^0_{{\phantom 0} 0}-T^r_{{\phantom r} r}=0$, then at best we would could only constrain $T^r_{{\phantom r} r}$ to be of the form $A/r^4$, where $A$ is a constant that decouples from (\ref{3.7y}). To see whether or not we could still secure $A=0$ anyway we need to impose the gravitational equations of motion given in (\ref{2.9x}) and (\ref{2.10x}), as this is the only further information involving $T^r_{{\phantom r} r}$ that is available. And as we shall, while it actually is then possible to secure $A=0$ it is not automatic, and one can  find solutions in which $A\neq 0$ as well.

\subsection{Solving the gravitational equations of motion}
\label{S3B}

To facilitate the evaluation of (\ref{3.2x}) and (\ref{3.3x}) we insert (\ref{3.4x}) into both of these two equations and obtain 
\begin{eqnarray}
T^r_{{\phantom r} r}=\frac{a^2B}{2}S^{4}-\frac{aB^{\prime}}{6}S^3-\frac{2aB}{3r}S^3+\frac{B^{\prime}}{6r}S^2+\frac{B}{6r^2}S^2 -\frac{1}{6r^2}S^2-\lambda S^4,
\label{3.9y}
\end{eqnarray}                                 
\begin{eqnarray}
2a^2BS^4-aB^{\prime}S^{3}-\frac{2aB}{r}S^{3}+\frac{S^2}{6}\left(B^{\prime\prime}+\frac{4B^{\prime}}{r}+\frac{2B}{r^2}-\frac{2}{r^2}\right)-4\lambda S^4=0.
\label{3.10y}
\end{eqnarray}                                 
With $S(r)=1/(b+ar)$, on  multiplying through by $S^{4}$ and conveniently by $6r^2$ we obtain
\begin{eqnarray}
6r^2S^{-4}T^r_{{\phantom r} r}=3a^2r^2B-ar^2B^{\prime}S^{-1}-4aBrS^{-1}+rB^{\prime}S^{-2}+BS^{-2} -S^{-2}-6r^2\lambda,
\label{3.11y}
\end{eqnarray}                                 
\begin{eqnarray}
12a^2r^2B-6ar^2B^{\prime}S^{-1}-12arBS^{-1}+S^{-2}\left(r^2B^{\prime\prime}+4rB^{\prime}+2B-2\right)-24r^2\lambda =0.
\label{3.12y}
\end{eqnarray}                                 
On inserting (\ref{2.18x}), on using (\ref{2.19x}) (but without setting  $T^r_{{\phantom r} r}=0$), and on setting $B(R_0<r<R_1)$ equal to a for the moment generic $w+u/r+vr-Kr^2$ (i.e., without regard to (\ref{2.16x}) and (\ref{2.17x})), for the respective $T^r_{\phantom{r}r}$ and scalar field equations we then obtain
\begin{align}
\frac{8\alpha_gS^{-4}}{r^2}(1+3uv-w^2)=
6r^2S^{-4}T^r_{{\phantom r} r}
=r^2\left(-abv-6\lambda-3Kb^2-a^2\right)
+r\left(2b^2v-2abw-2ab\right)+wb^2-b^2-3abu,
\label{3.13y}
\end{align}                                 
\begin{eqnarray}
r^2\left(-6abv-24\lambda-12Kb^2+2a^2w-2a^2\right)
+r\left(6b^2v-8abw-4ab+6a^2u\right)+2wb^2-2b^2-6abu
=0.
\label{3.14y}
\end{eqnarray}                                 
From the need to satisfy the scalar field equation given in (\ref{3.14y}) for all $r$ it follows that 
\begin{eqnarray}
-3abv-12\lambda-6Kb^2+a^2w-a^2=0,\quad 
3b^2v-4abw-2ab+3a^2u=0,\quad wb^2-b^2-3abu=0.
\label{3.15y}
\end{eqnarray}                                 
We note that (\ref{3.15y}) does not force $w^2-1-3uv$ to vanish. On inserting (\ref{3.15y})  into the $T^r_{\phantom{r}r}$ equation given in (\ref{3.13y}) we obtain 
\begin{eqnarray}
\frac{8\alpha_g}{r^2}(1+3uv-w^2)=\frac{1}{(b+ar)^4}\left[
\frac{r^2}{2}\left(abv-a^2-a^2w\right)
+\frac{2ar}{3}\left(bw-b-3au\right)\right].
\label{3.16y}
\end{eqnarray}                                 
Noting now that  for general $a$ and $b$ the two sides of (\ref{3.16y}) have a totally different dependence on $r$, from (\ref{3.16y}) we can conclude that both sides of (\ref{3.16y}) must vanish separately, i.e., that  $T^r_{{\phantom r} r}$ and  thus $W^r_{{\phantom r} r}$ must vanish after all. We thus obtain 
\begin{eqnarray}
1+3uv-w^2=0\quad bw-b-3au=0,\quad abv-a^2-a^2w=0,
\label{3.17y}
\end{eqnarray}                                 
so that now we can set $w^2-1-3uv=0$.
The latter two  relations given in (\ref{3.17y}) are compatible  with those given in (\ref{3.15y}). And thus for general $a$ and $b$ there is a consistent overall completely exact solution of the form 
\begin{eqnarray}
w^2=1+3uv,\quad b(w-1)=3au,\quad b^2v=2ab+3a^2u, \quad 6\lambda+3Kb^2=-2a^2-a^2w.
\label{3.18y}
\end{eqnarray}                                 
(The third equations given in (\ref{3.17y}) and (\ref{3.18y}) are not independent of the first two,  but for convenience we list them.) With (\ref{3.18y}) we thus recover the result assumed but not apparently derived in  \cite{Brihaye2009,Horne2016,Hobson2021} that $w^2=1+3uv$.

\subsection{A special case: $a=0$}
\label{S3C}

If we set $a=0$ so that $S(r)=1/b$, then as noted in \cite{Horne2016},  we obtain a metric with no linear term, viz. 
\begin{eqnarray}
w=1,\quad v=0,  \quad 2\lambda+Kb^2=0,\quad B(r)=1+\frac{u}{r}-Kr^2,
\label{3.19z}
\end{eqnarray}                                 
with $u$ undetermined. The reason why $u$ is undetermined is because if we insert $B=1+u/r$ into the form for the Ricci scalar given in (\ref{2.14x}) we would find that it drops out identically (as it would of course have to do since the Schwarzschild solution is Ricci flat).

The emergence of the structure given in  (\ref{3.19z}) is familiar. Specifically, in \cite{Mannheim1991}  it was noted that one can transform the metric $B(r)=1/A(r)=w+u/r+vr-Kr^2$ with a linear term into a metric of the form $b(\rho)=1/a(\rho)=w^{\prime}+u^{\prime}/\rho-K^{\prime}\rho^2$ where there is no linear term. The procedure is to set $a(r)=1/b(r)$ in (\ref{2.4x}), so that (\ref{2.4x}) then integrates to
\begin{eqnarray}
\frac{p(r)}{r}=\frac{1}{1+cr},
\label{3.20z}
\end{eqnarray}                                 
where $c$ is an integration constant. On setting $\rho=p(r)=r/(1+cr)$, i.e., $r=\rho/(1-c\rho )$, (\ref{2.5x}) takes the form of (\ref{2.2x}) with 
\begin{eqnarray}
b(\rho)=\frac{1}{a(\rho)}=w-3uc+\frac{u}{\rho}+\rho(v-2wc+3uc^2)-\rho^2(K+vc+uc^3-wc^2).
\label{3.21z}
\end{eqnarray}
On now setting $c=a/b$, through use of (\ref{3.18y}) we find that  (\ref{3.21z}) takes the form
\begin{eqnarray}
b(\rho)=\frac{1}{a(\rho)}=1+\frac{u}{\rho}+\frac{2\lambda\rho^2}{b^2},
\label{3.22z}
\end{eqnarray}
to thus be of the same form as the Schwarzschild de Sitter solution of standard gravity. (However, as we discuss in Sec. \ref{S5}, the allowed range for $\rho$ is constrained and differs from the range for $\rho$ that is required in the Schwarzschild de Sitter solution.) Then, with $c=a/b$  the conformal factor in (\ref{2.5x}) has the form $p(r)/r=b/(b+ar)$. Up to an overall constant we recognize this form for $p(r)/r$ as being none other than the form for $S(r)$  given in (\ref{3.4x}). With the scalar field transforming as $e^{-\alpha}S=rS/p(r)$ under a conformal transformation, as noted in \cite{Hobson2021} this form for $p(r)/r$ precisely brings $S(r)$ to the constant value $S(r)=1/b$. We discuss the significance of this result in Sec. \ref{S5}.

In addition, we would like to note that the our initial derivation of the form for the metric given in (\ref{2.6x}) was a purely kinematic derivation, with the $A(r)=1/B(r)$ relation holding without needing to impose the gravitational equations of motion. And as such, this relation holds in both of the $T_{\mu\nu}=0$ and $T_{\mu\nu}\neq 0$ regions. However, the identification of $a(\rho)$ with $1/b(\rho)$ in (\ref{3.22z}) only holds in the $T_{\mu\nu}=0$ region.  And not only that, it only holds for the particular metric that satisfies the exterior region gravitational equations of motion. The $a(\rho)=1/b(\rho)$ relation does not hold in the $T_{\mu\nu}\neq 0$ region even though the $A(r)=1/B(r)$ relation does. Consequently, unlike in the $T_{\mu\nu}=0$ region, there cannot be any conformal equivalence between solutions in the $T_{\mu\nu}\neq 0$ region. 

 \subsection{A special case: $b=0$}
\label{S3D}

While the solutions described above lead to $T^r_{{\phantom r} r}=0$, there is another solution in which $T^r_{{\phantom r} r}$ is not forced to be zero at all. Specifically, if we set $b=0$ in (\ref{3.13y}) and (\ref{3.14y}) we obtain
\begin{align}
\frac{8\alpha_ga^4r^4}{r^2}(1+3uv-w^2)=
6a^4r^6T^r_{{\phantom r} r}
=r^2\left(-6\lambda-a^2\right),\quad 
r^2\left(-24\lambda+2a^2w-2a^2\right)
+6ra^2u
=0,
\label{3.23z}
\end{align}                                 
with solution
\begin{eqnarray}
u=0,\quad 12\lambda=a^2(w-1),\quad 8\alpha_ga^4(w^2-1)=6\lambda +a^2.
\label{3.24z}
\end{eqnarray}
Then, with $w+1\neq 0$ we obtain
\begin{eqnarray}
16a^2\alpha_g(w-1)=1, \quad 12\lambda=\frac{1}{16\alpha_g}.
\label{3.25z}
\end{eqnarray}
As we see, just as $a=0$ leads to $v=0$, similarly $b=0$ leads to $u=0$. Now while this may not be particularly interesting physically it does show that in principle setting $f(r)=0$ does not force $T^r_{{\phantom r} r}$ to be zero, and not even after we impose the equations of motion. As we had noted in (\ref{3.7y}), setting $T^0_{{\phantom 0} 0}-T^r_{{\phantom r} r}=0$ does not fix the coefficient of the $1/r^4$ term in $T^r_{{\phantom r} r}$. Similarly, setting $W^0_{{\phantom 0} 0}-W^r_{{\phantom r} r}=0$ does not fix the coefficient of the $1/r^4$ term in $W^r_{{\phantom r} r}$. The $b=0$ solution leads to both $W^r_{{\phantom r} r}$ and $T^r_{{\phantom r} r}$ behaving as $1/r^4$, and neither of them has to separately vanish. There is thus no general theorem that the vanishing of $T^0_{{\phantom 0} 0}-T^r_{{\phantom r} r}$ requires the vanishing of $T^r_{{\phantom r} r}$, and not even if we impose the gravitational equations of motion. Having now established the general implications of setting $T^0_{{\phantom 0} 0}-T^r_{{\phantom r} r}=0$, in Sec. \ref{S4}  we will study geodesics in the most interesting case, namely where $T^r_{{\phantom r} r}$ is zero, i.e., we consider $S(r)=1/(b+ar)$ with neither $a$ nor $b$ set equal to zero. In order to do this we first need to relate $a$ and $b$ to the metric coefficients.

\subsection{Fixing $a$ and $b$}
\label{S3E}

For the general case given in (\ref{3.18y}) we can determine $a$ and $b$ in terms of the parameters that appear in $B(r)=w+u/r+vr-Kr^2$, parameters that are determined in  (\ref{2.17x}) by  dynamical relations that do not involve either $a$ or $b$ at all. (With $T^0_{{\phantom 0} 0}-T^r_{{\phantom r} r}$ vanishing if $S(r)=1/(b+ar)$, the general solution to $W^0_{{\phantom 0} 0}-W^r_{{\phantom r} r}=0$ is then independent of $a$ or $b$; with $a$ and $b$ then only being fixed in terms of $(w,u,v,K)$ by the $4\alpha_gW^r_{{\phantom r} r}=T^r_{{\phantom r} r}$ and scalar field equations.) 

From the $w^2=1+3uv$ and  $b(w-1)=3ua$ relations given in  (\ref{3.18y}) we obtain
\begin{eqnarray}
\frac{a}{b}=\frac{v}{w+1}.
\label{3.26z}
\end{eqnarray}
Then from the $6\lambda+3Kb^2=-2a^2-a^2w$ relation given in  (\ref{3.18y}) we obtain 
\begin{eqnarray}
b^2\left(3K+\frac{(w+2)v^2}{(w+1)^2}\right)+6\lambda=0.
\label{3.27z}
\end{eqnarray}
With the quantity $uv=-2N^*\beta^*(N^*\gamma^*+\gamma_0)$ for a galaxy with   $N^*$ stars being found to be very much less than one in the galactic rotation curves fits given in \cite{Mannheim2011,Mannheim2012,O'Brien2012}, and with $w$ thus being close to one, we can set 
\begin{eqnarray}
 \frac{a}{b}= \frac{v}{2},\quad b^2=-\frac{8\lambda}{v^2+4K}.
\label{3.28z}
\end{eqnarray}
Consequently, we can write $S(r)$ as
\begin{eqnarray}
S(r)&=&\frac{(v^2+4K)^{1/2}}{2(-2\lambda)^{1/2}(1+vr/2)}
=\frac{(v^2+4K)^{1/2}}{2(-2\lambda)^{1/2}}\left(1-\frac{vr}{2}\right)
\nonumber\\
&=&\frac{[(N^*\gamma^*+\gamma_0)^2+4(\kappa^*+K_0)]^{1/2}}{2(-2\lambda)^{1/2}}\left(1-\frac{(N^*\gamma^*+\gamma_0)r}{2}\right),
\label{3.29z}
\end{eqnarray}
with the last two steps following since $vr \ll 1$ in the galactic regime of interest.
Thus in this regime we can set
\begin{eqnarray}
\frac{rS^{\prime}(r)}{S(r)}=-\frac{vr}{2}=-\frac{(N^*\gamma^*+\gamma_0)r}{2},
\label{3.30z}
\end{eqnarray}
a quantity that we shall meet in Sec. \ref{S4A}. Thus we see that both for $S(r)$ and $rS^{\prime}(r)/S(r)$ the contribution to $S(r>R_0)$ due to the sources in $r<R_0$ is of order $N^*$. This will prove to be very consequential in the following.

\section{Implication for Geodesics}
\label{S4}

\subsection{The concern raised by Horne}
\label{S4A}

While a test particle action of the standard $I_T=-m\int ds$ form is not conformal invariant, one can construct a test particle action that is, viz. \cite{Mannheim1993}
\begin{eqnarray}
I_T=-h\int S(x)ds,
\label{4.1y}
\end{eqnarray}
where $S(x)$ is a scalar field. The variation of this action with respect to the particle coordinate $x^{\lambda}$ leads to the trajectory \cite{Mannheim1993}
\begin{eqnarray}
\frac{d^2x^{\lambda} }{ ds^2}
+\Gamma^{\lambda}_{\mu \nu} 
\frac{dx^{\mu}}{ ds}\frac{dx^{\nu } }{ds}=
-\frac{1}{S}
\left[g^{\lambda \mu}+\frac{dx^{\lambda}}{ds}\frac{dx^{\mu}}{ds}\right]\frac{\partial S}{\partial x^{\mu}}.
\label{4.2y}
\end{eqnarray}
This trajectory only reduces to 
\begin{eqnarray}
\frac{d^2x^{\lambda} }{ ds^2}
+\Gamma^{\lambda}_{\mu \nu} 
\frac{dx^{\mu}}{ ds}\frac{dx^{\nu } }{ds}=0
\label{4.3y}
\end{eqnarray}
if $S(x)$ is constant. In \cite{Mannheim2012} it was shown that in a general metric of the form $ds^2=B(r)dt^2-A(r)dr^2-r^2d\theta^2-r^2\sin^2\theta d\phi^2$ trajectories that obey the scalar-field-dependent (\ref{4.2y}) with a radially-dependent $S(r)$ are of the form
\begin{eqnarray}
c\frac{d^2t}{ds^2}+\frac{cB^{\prime}}{B}\frac{dt}{ds}\frac{dr}{ds}
&=&-\frac{cS^{\prime}}{S}\frac{dt}{ds}\frac{dr}{ds},
\nonumber \\
\frac{d^2r}{ds^2}+\frac{A^{\prime}}{2A}\left(\frac{dr}{ds}\right)^2
-\frac{r}{A}\left(\frac{d\theta}{ds}\right)^2
- \frac{r{\rm sin}^2\theta}{A}\left(\frac{d\phi}{ds}\right)^2
+\frac{c^2B^{\prime}}{2A}\left(\frac{dt}{ds}\right)^2
&=&-\frac{S^{\prime}}{AS} 
-\frac{S^{\prime}}{S}\left(\frac{dr}{ds}\right)^2,
\nonumber \\
\frac{d^2\theta}{ds^2}
+\frac{2}{r}\frac{d\theta}{ds}\frac{dr}{ds}
-{\rm sin}\theta{\rm cos}\theta\left(\frac{d\phi}{ds}\right)^2
&=&-\frac{S^{\prime}}{S}\frac{d\theta}{ds}\frac{dr}{ds},
\nonumber \\
\frac{d^2\phi}{ds^2}
+\frac{2}{r}\frac{d\phi}{ds}\frac{dr}{ds}
+2\frac{{\rm cos}\theta}{{\rm sin}\theta}\frac{d\phi}{ds}\frac{d\theta}{ds}
&=&-\frac{S^{\prime}}{S}\frac{d\phi}{ds}\frac{dr}{ds},
\label{4.4yz}
\end{eqnarray}                                 
with the prime denoting differentiation with respect to $r$. Thus for circular orbits with  $dr/ds=0$, $d\theta/ds=0$, $dt/ds=1/c$, $s=ct$, $\theta=\pi/2$  we obtain  
\begin{eqnarray}
\frac{r^2}{c^2}\left(\frac{d\phi}{dt}\right)^2=\frac{rB^{\prime}}{2}+\frac{rS^{\prime}}{S},
\label{4.5y}
\end{eqnarray}
regardless of whether or not $A=1/B$ \cite{footnoteE}. These orbits thus depart from the standard $(r^2/c^2)(d\phi/dt)^2=rB^{\prime}/2$ that would be associated with (\ref{4.3y}). Now the successful fitting to the rotation curves of 138 spiral galaxies using (\ref{2.18x}) with universal parameters and no dark matter (or its 276 free galactic dark matter halo parameters for the 138 galaxy sample) that was reported in \cite{Mannheim2012,Mannheim2011,O'Brien2012} relied heavily on the linear potential term given in (\ref{2.18x}). (Flat rotation curves come about through a balance between the falling $u/r$ and the rising $vr$.) However, the analysis given above indicates that to support (\ref{2.18x}) we need a scalar field $S(r)=1/(b+ar)$. Thus if we are to use (\ref{2.18x}) we should use (\ref{4.5y}) rather than the standard $(r^2/c^2)(d\phi/dt)^2=rB^{\prime}/2$ that was used in \cite{Mannheim2012,Mannheim2011,O'Brien2012}. This then calls into question \cite{Horne2016} the fitting given in \cite{Mannheim2012,Mannheim2011,O'Brien2012}.

To see the difficulty in detail, we note that from $S(r)=1/(b+ar)$ and from (\ref{3.18y}) we obtain 
\begin{eqnarray}
\frac{rS^{\prime}}{S}=-\frac{ar}{b+ar},\quad \frac{a}{b}=\frac{(w-1)}{3u}=\frac{v}{w+1}.
\label{4.6yy}
\end{eqnarray}
Now according to (\ref{2.21x})  we can set $B=w+u/r+vr-Kr^2$  where $w=(1+3uv)^{1/2}$. For weak gravity the fits of \cite{Mannheim2012,Mannheim2011,O'Brien2012} lead to $uv\ll 1$ in galactic regions where $u/r$, $vr$ and $-Kr^2$ are all very much smaller than one. Consequently, with $w\approx 1$ and $w+1\gg vr$ (\ref{4.5y}) takes the form 
\begin{eqnarray}
\frac{r^2}{c^2}\left(\frac{d\phi}{dt}\right)^2=-\frac{u}{2r}+\frac{vr}{2}-Kr^2-\frac{vr}{w+1+vr}=-\frac{u}{2r}+\frac{vr}{2}-Kr^2-\frac{vr}{2}=-\frac{u}{2r}-Kr^2.
\label{4.7yy}
\end{eqnarray}
Thus as noted in \cite{Horne2016}, the linear term drops out identically. Moreover, if we instead use the solution with $a=0$ (so that we then could  use the standard geodesic given in (\ref{4.3y})), then according to (\ref{4.6yy}) $v$ is then zero identically. Thus the linear term is lost whether we work with $S(r)=1/(b+ar)$ or with $S(r)=1/b$. With the linear term being central to the conformal gravity fitting to galactic rotation curves this then is the concern raised by Horne and reemphasized by Hobson and Lasenby. 

\subsection{Shortcomings of  the macroscopic scalar field case}
\label{S4B}  

To address the concern identified in (\ref{4.7yy})  we note that the linear potential contribution to the $rS^{\prime}/S$ term in (\ref{4.5y}) is only able to cancel the  linear potential contribution to the $rB^{\prime}/2$ term in (\ref{4.5y}) because these two contributions are equal. Now according to (\ref{2.17x}), for a single star in the cosmological background  the linear potential $v$ term is given by $\gamma_0+\gamma^*$, and thus for a galaxy with $N^*$ stars  the $v$ term  is given by $\gamma_0+N^*\gamma^*$. Hence the effect of the $I_T=-h\int S(x)ds$ test particle action is to not only give the test particle a mass that is as big as both the cosmological contribution and the number of stars in the galaxy around which it is orbiting, by depending on $N^*$ it gives the test particle a mass that even  changes with the environment in which it propagates. In fact, if we set $S=1/(b+ar)$, then from (\ref{3.28z}) and (\ref{2.17x}) we see that
\begin{eqnarray}
\frac{1}{b}=\left(\frac{(\gamma_0+N^*\gamma^*)^2+4K_0+4\kappa^*}{-8\lambda}\right)^{1/2}.
\label{4.8w}
\end{eqnarray}
Thus even if we ignore the cosmological background,  $1/b$ is given by 
\begin{eqnarray}
\frac{1}{b}=\left(\frac{(N^*\gamma^*)^2+4\kappa^*}{-8\lambda}\right)^{1/2},
\label{4.9w}
\end{eqnarray}
to not only make the magnitude of $S(r)$ (whether it be $1/(b+ar)$ or $1/b$) be of the order of the number of stars $N^*$ located in the $r<R_0$ region, but to even be affected by the contribution of clusters of galaxies in the $r>R_1$ region.

Since the analysis given above would lead to a galactic contribution of order $N^*$ and a cosmological contribution of order $\gamma_0$ to the test particle orbit we have to conclude that  if there is a macroscopic scalar field,  the test particle action $I_T=-\int S(x)ds$ could not describe the mass generation mechanism for the test particle. Rather, just like the way $e\int dx^{\lambda}A_{\lambda}$ describes a particle propagating in an external long range electromagnetic field, the $I_T=-\int S(x)ds$ action describes a particle propagating in a macroscopic long range scalar field, should one exist. For mass generation we thus have to look elsewhere, something we will do in Sec. \ref{S6} below. 

\subsection{Inconsistencies in the pure scalar field calculation}
\label{S4C}

In regard to the analysis that led us to (\ref{4.7yy}), we note that  is actually internally inconsistent. Specifically, in order for the test particle and the scalar field to couple to each other so as to provide us with the  $I_T=-\int S(x)ds$ action in the first place, we would need some interaction between them. If we take  the fermion to obey the the Yukawa coupled  (\ref{1.7y}), then as  shown in \cite{Mannheim2021}, in the eikonal approximation the fermion will move on the trajectory described by (\ref{4.2y}), viz. the one associated with the test particle action given in (\ref{4.1y}). Thus, the use of (\ref{4.2y}) requires the existence of a long range scalar field that appears in the $hS\psi$ term in the fermion wave equation. However, its presence in the fermion wave equation requires the presence of the $h\bar{\psi}\psi$ term in the scalar field wave equation given in (\ref{1.8y}). But then we could not use the fermion-independent scalar field wave equation
given in (\ref{3.3x})) that we did use in order to establish that we could set $S(r)=1/(a+br)$ in the first place.  Thus if the fermion-scalar coupling is of significance we could not obtain (\ref{4.7yy}). Thus we could only recover (\ref{4.7yy}) if the coupling term is not significant, in which case the $rB^{\prime}/2$ term would then dominate in (\ref{4.5y}) and the geodesics would be standard.

A further shortcoming of the analysis stems from the fact that the test particle of relevance to galactic rotation curves is itself a star. Thus there will not just be a coupling of the form $hS\bar{\psi}\psi$ for the test particle, the test particle  will have energy and momentum of its own. Thus, as per (\ref{1.9y}), it will contribute a term of the form
\begin{eqnarray}
T^{\mu \nu}=i\bar{\psi}\gamma^{c}V_{\mu c}(x)[\partial_{\nu}+\Gamma_{\nu}(x)]\psi
\label{4.12yy}
\end{eqnarray}
in the $R_0<r<R_1$ region that we are considering. So again the analysis of just a scalar field $S(x)$ in this region is invalid. Now while this particular modification to the analysis would appear be overwhelmed by the $N^*\gamma^*$ and $\gamma_0$ contributions to $B(r)$, the fermion obeys the Yukawa coupled (\ref{1.7y}), and with $S$ being of order $N^*$, solutions to the Dirac equation would be of order $N^*$ too. 

Because of its spherical symmetry the analysis that leads to (\ref{4.7yy}) only applies to a spherical matter source such a single star and not to a spiral galaxy. And this creates an additional difficulty.  Specifically, absent any long range scalar field, the standard approach to weak gravity is to first treat each star as an isolated spherical system and solve for the metric exterior to it. Then, all the ensuing potentials are added up using linear superposition. However, if there is a scalar field in the exterior region we could not use linear superposition unless each star were to have its own independent scalar field. But there is only one scalar field, just as there is only one gravitational field. Thus for more than one star we would need to solve for a scalar field in the presence of many sources, sources that  for a spiral galaxy are not distributed spherically. This is not the calculation provided in Sec. \ref{S3}. In Sec. \ref{S5} we discuss this many-source issue in the weak gravity limit.

An even more problematic concern in having a long range, massless scalar field may be seen by considering a  massless photon propagating in the $r>R_0$ region. Taking the only source to be the $f(r)$ source in the $r<R_0$ region, then since the photon does not couple to the scalar field it will travel on a null geodesic associated with $B^*(r)$. Moreover, a massless fermion would equally travel on the same null geodesic. Then if we give the fermion a mass, its trajectory would not be modified very much. Moreover, if we take the test particle to be a star,  compared with a whole galaxy (let alone the cosmology of the entire Universe), then to one part in $10^{11}$ its orbit should not differ from that of an $S$-independent geodesic. (It was precisely this  argument that was presented in \cite{Mannheim2012}). Thus whatever is to happen to the test particle we should not expect it to be an effect of order $N^*$. So again, a macroscopic scalar field could not describe mass generation. Below in Sec. \ref{S6} we show that an effect of order $N^*$ does not appear if the mass generation is microscopic, since then $S(x)$ is of order the mass of the test particle. 

Moreover, if the fermion were to have a mass such that its coupling to the geometry would be of the form $I_T=-m\int ds$, it would follow the standard geodesics and give rise to orbital velocities of the form $r^2(d\phi/dt)^2/c^2=rB^{\prime}/2$. (If $S(x)$ is equal to a constant $m$ there would be no $rS^{\prime}(r)/S(r)$ term in (\ref{4.5y}).) If the test particle is to get its mass via $I_T=-\int S(x)ds$ instead, we should not expect this to lead to a radical departure from $r^2(d\phi/dt)^2/c^2=rB^{\prime}/2$. And yet if there is a macroscopic mass-generating field it does. Hence, there must be no long range, macroscopic scalar field to which the test particle could couple to in the first place, and we must thus dispense with such long range mass-generating scalar fields. Thus we must also dispense with a test particle action of the form $I_T=-\int S(x)ds$ and replace it by $I_T=-m\int ds$. While this latter action is not conformal invariant, nonetheless in Sec. \ref{S6} we shall show that in fact this is the test particle action that  actually is relevant, as it is associated with the dynamical symmetry breaking mean-field approximation to a four-fermion theory as associated with critical scaling and anomalous dimensions, with the change in vacuum from normal to broken producing a mean-field approximation that is not  conformal invariant.

\section{Conformal Transformation of the Brihaye-Verbin Solution}
\label{S5}

In regard to long range, macroscopic scalar fields there is one other issue we wish to address, namely the degree to which the $S(r)=1/(b+ar)$ and $S=1/b$ scalar field solutions can be related.  As well as study the solution with $S(r)=1/(b+ar)$, we can follow \cite{Horne2016} and \cite{Hobson2021} and study the solution in which $a=0$. To make contact with this solution  we can make coordinate and conformal transformations on the $B(r)=w+u/r+vr-Kr^2$ and $S(r)=1/(b+ar)$ solution that we obtained above. Thus if we now make the conformal and coordinate transformations described in Sec. \ref{S3C}, then with $r=\rho/(1-c\rho)$, $\rho_0=R_0/(1+cR_0)$, $\rho_1=R_1/(1+cR_1)$ and $c=a/b$ we obtain 
\begin{eqnarray}
b(\rho_0<\rho<\rho_1)=\frac{1}{a(\rho_0<\rho<\rho_1)}=1+\frac{u}{\rho}+\frac{2\lambda\rho^2}{b^2}.
\label{5.1z}
\end{eqnarray}                                 
Thus, as shown directly in  \cite{Horne2016} by solving the equations of motion with $S(r)=1/b$, the linear term drops out identically.

While this metric has the same generic from as a Schwarzschild de Sitter metric, it only does so in the $R_0<r<R_1$ region. In the $r>R_1$ region (viz. $\rho>R_1/(1+cR_1)=\rho_1$) we would be sensitive to the $-\tfrac{1}{2}\int_r^{\infty} dr^{\prime}r^{\prime 3}f(r^{\prime})-
(r^2/6)\int_r^{\infty}
dr^{\prime}r^{\prime }f(r^{\prime})$  term in (\ref{2.11x}), and so in $r>R_1$ the same coordinate and conformal transformations would lead us to 
\begin{eqnarray}
b(\rho_0<\rho)&=&(1-c\rho )^2B(R_0<r)=1+\frac{u}{\rho}+\frac{2\lambda \rho^2}{b^2}-\frac{(1-c\rho)^2}{2}\int_{\rho/(1-c \rho)}^{\infty}
dr^{\prime}r^{\prime 3}f(r^{\prime})
\nonumber\\
&-&\frac{\rho^2}{6}\int_{\rho/(1-c \rho)}^{\infty}
dr^{\prime}r^{\prime }f(r^{\prime})
\label{5.2z}
\end{eqnarray}                                 
instead. Since $b(\rho)$ would have to behave as $b(\rho)=1+u/\rho +2\lambda  \rho^2/b^2$ for all allowed values of $\rho$ in order to be a Schwarzschild de Sitter metric, the metric given in (\ref{5.2z}) is not of this form and can thus not be identified as a Schwarzschild de Sitter metric.

Now we had noted  that the exterior $S(r)=1/(b+ar)$ and $S(r)=1/b$ solutions were conformally and coordinate equivalent. However, we now note that  the $r=\rho/(1-c\rho )$ transformation is singular at $\rho=1/c$ unless $c$ is negative, in which case the $\rho/r=p(r)/r=1/(1+cr)$ conformal transformation is singular ($p(r)$ being given in (\ref{2.4x})). Moreover, while $r=0$ and $\rho=0$ coincide, if $c>0$ then $r\rightarrow \infty$ at $\rho=1/c$, while $\rho=r/(1+cr)$ is bounded between $0$ and $1/c$. Now a reader might initially object that because of the $-Kr^2$ term the range for $r$ actually is bounded. However that is only the case if $K$ is positive (though even then the bound is of order $r=1/K^{1/2}$, $\rho=1/(c+K^{1/2})$, i.e., not $\rho=1/c$), and the transformation between the $S(r)=1/(b+ar)$ and the $S(r)=1/b$ sectors would anyway be the same even if $K$ is negative or zero, i.e., even if $r$ is unbounded. With $\rho=r/(1+cr)$ and $r=\rho/(1-c\rho )$ we see that $r$ is bounded if $c<0$ while $\rho$ is bounded if $c>0$. 

Now  a pure Schwarzschild geometry (viz. (\ref{5.1z}) with $\lambda=0$) would require the range of the radial coordinate to be unbounded, and yet there is no choice for the sign of $c$ for which $r$ and $\rho$ could both be unbounded. Thus in this regard too we cannot regard (\ref{5.1z}) with $\lambda=0$ as being a pure Schwarzschild metric. Moreover, as had been noted in \cite{Mannheim1991}, for either sign of $c$ one of the $r\rightarrow\rho/(1-c\rho)$ and $S\rightarrow (1+cr)S$ transformations is singular, with the $S(r)=1/(b+ar)$ and $S(r)=1/b$ solutions belonging to conformal sectors that are separate and distinct from one another. Moreover, since the fits of \cite{Mannheim2012,Mannheim2011,O'Brien2012} yield $u$ negative and $v$ positive, the fits give $c=a/b>0$. Then, since we want to work in a non-compact space in which the range of the radial coordinate is infinite (as must be the case if the global cosmological spatial three-curvature $k$ is negative, with it being shown in \cite{Mannheim1989,Mannheim2006} and the appendix that in conformal cosmology $k$ is negative and  given by the negative $k=-\gamma_0^2/4-K_0$), we must thus work in the sector in which $S(r)=1/(b+ar)$. I.e., we must work in the sector in which there is a linear potential term $vr$, and thus use $B(R_0<r)$ rather than $b(\rho_0<\rho)$ for galactic  rotation curve fits (regardless in fact of whether or not we are working with a long range scalar field). The $B(R_0<r)$ sector is actually the most general sector of the theory since in it both the second- and fourth-moment integrals of the local inhomogeneous source $f(r)$ couple to the metric as per (\ref{2.17x}). Thus before making any transformation (either with or without any scalar field $S(x)$) we should start with the most general form for the metric, viz. the most general solution to $W_{\mu\nu}=0$ as given in (\ref{2.16x}).

Now in making the transformation from the metric with a $vr$ term to the one without we relied on radial symmetry. However in a spiral galaxy there are $N^*$ identical sources and they are not distributed spherically. For the galaxy the weak gravity metric due to localized sources (each with an $f(r)$ that is only nonzero in $r<R_0^i$)  is given by 
\begin{eqnarray}
B(r)=\sum_i \left(w+\frac{u}{|\vec{r}-\vec{r}_i|}+v|\vec{r}-\vec{r}_i|\right).
\label{5.4z}
\end{eqnarray}
Now unlike the single source case this time $B(r)$ depends not just on the $ r-r_i$ radial magnitude  but also on the angle between $\vec{r}$ and $\vec{r}_i$ as the galactic disk is not spherically symmetric, with $|\vec{r}-\vec{r}_i|=(r^2+r_i^2-2rr_i\cos\phi_i)^{1/2}$. Thus on transforming to $b(\rho)$ we obtain 
\begin{eqnarray}
b(\rho)=(1-c\rho)^2B(r)=(1-c\rho)^2\sum_i \left(w+\frac{u}{X_i}+vX_i\right),
\label{5.5z}
\end{eqnarray}
where
\begin{eqnarray}
X_i=\frac{[\rho^2(1-c\rho_i)^2+\rho_i^2(1-c\rho)^2-2\rho\rho_i(1-c\rho)(1-c\rho_i)\cos\phi_i]^{1/2}}{(1-c\rho)(1-c\rho_i)}.
\label{5.6z}
\end{eqnarray}
As we see, the $v$-dependent term has not in fact been cancelled, since we are only allowed one overall coordinate transformation on all the coordinates and not separate ones for each individual star. Thus as we see, while for a single star we could remove the term that is linear in $\rho$, by coordinate and conformal transformations  we could not do so for an entire spiral galaxy, as we are only allowed one  overall coordinate transformation and one overall conformal transformation.

Finally, we should note that even if we use the metric given (\ref{5.1z}) and set $S=1/b$, then while the constancy of $S$ now leads to $S$-independent 
geodesic test particle motion, according to (\ref{4.8w}) the mass of the test particle would still be of order $N^*$, and thus still be unacceptable.
 
\section{Dynamical Symmetry Breaking and a Microscopic Scalar Field}
\label{S6}

To conclude this paper we turn to the role played by scalar fields in microscopic, quantum-field-theoretic mass generation. Such mass generation has  two key features that distinguish it from the macroscopic study that we presented above. The first  is that relevant scalar fields  only have a spatial variation within the particles to which they give mass. And the second is that in the broken vacuum the action is no longer conformal invariant.

\begin{figure}[htpb!]
  \centering
   \includegraphics[scale=1.0]{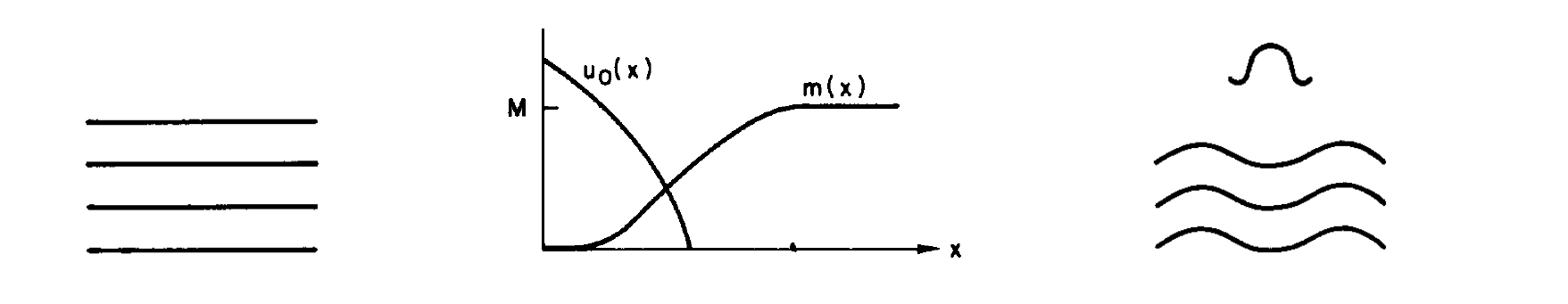}
  \caption{undistorted negative energy states, bound first positive energy state, distorted states}
   \label{bag}
\end{figure}

For the first issue we note that prior to the early universe electroweak symmetry breaking phase transition the gravitational source consists of massless fermions and massless gauge bosons. Together they provide the radiation fluid that leads to the background cosmological Robertson-Walker geometry. With the fermions being massless they cannot come to rest and form bound systems such as stars, and they thus have to have wave functions that fill all space. (With $k=0$ these modes are massless plane waves, for $k<0$ they are associated Legendre functions, and for $k>0$ they are Gegenbauer polynomials.) Following the phase transition that occurs as the Universe cools down the fermions then do acquire masses by the electroweak symmetry breaking mechanism. Our actual interest is in this breaking being done by fermion bilinears (see \cite{Mannheim2017} for a detailed analysis). For the Dirac equation the vacuum $|\Omega\rangle$ consists of a filled negative energy sea so that $S(x)=\langle \Omega |\bar{\psi}\psi|\Omega \rangle$ is a constant. As shown in the left panel in Fig. \ref{bag}, in this vacuum the negative energy modes still permeate all space only now as massive plane waves or their $k\neq 0$ analogs. With $\langle \Omega |\bar{\psi}\psi|\Omega \rangle$  being a constant it just contributes an additional constant value to the cosmological constant, and at this point the geometry is still homogeneous and isotropic.

%1

Given this vacuum structure it would thus be expected that the lowest lying positive energy mode would also be a massive plane wave, and that the geometry would remain homogeneous and isotropic with a vanishing Weyl tensor. However, it turns out that there is a positive energy state that lies lower in energy than a massive plane wave. Specifically, as noted in \cite{Dashen1974,Bardeen1975,Dashen1975}, in this state, known as a coherent state $|C\rangle$,  the scalar field takes a spatially-dependent expectation value $S(x)=\langle C |\bar{\psi}(x)\psi(x)|C \rangle$. As shown in the middle panel in Fig. \ref{bag} with $m(x)$ denoting $S(x)$ this expectation value is constant everywhere except in a small region where it forms a potential well. Now on its own this state is not stable. However, in the well that is formed one can bind the lowest lying positive energy fermion with wave function $u_0(x)=\langle x|b^{\dagger}|C\rangle$, with the entire configuration (coherent state plus fermion) being stable, where the creation operator $b^{\dagger}$ is the conjugate of the annihilation operator $b$ that effects $b|C\rangle=0$. As shown in the right panel in Fig. \ref{bag} the negative energy wave functions are also slightly modified in this region. However, they still take their prior plane wave form asymptotically. In the well in which it is bound the positive energy fermion wave function has both a height and a width, just as needed in order to support both of the moment integrals given in (\ref{2.17x}), with the second moment being due to the height [$f(r)$ can contain a delta function], and the fourth moment being due to the width [$f(r)$ needing to contain a derivative of a delta function]. The width can be understood as the radius of the fermion, and since there is a non-vanishing height the contribution of the fermion to the second-moment integral, and thus to the parameter $v$, cannot be zero. The dynamically induced scalar field thus plays two roles: it both generates fermion masses and provides a mechanism for fermions to localize. These localized fermions thus provide inhomogeneities in the homogeneous cosmological background, with this localization leading to a nonvanishing Weyl tensor. With the scalar field approaching its asymptotic background value exponentially fast \cite{Dashen1975,Mannheim1976}, it is only varying within the localized source, with its contribution to the energy-momentum tensor outside of the source being in the form of a cosmological constant. In consequence, outside the extended source the geometry receives no contributions from any spatial variation in  $S(x)=\langle C |\bar{\psi}(x)\psi(x)|C \rangle$.

In regard to the second feature, we recall  the familiar double-well potential $V(\phi)=-\mu^2\phi^2/2+\lambda\phi^4/4$, where $\phi$ is a quantum field. In the tree approximation $\phi$ acquires a vacuum expectation value $S(x)=\langle \Omega|\phi|\Omega \rangle=\mu/\lambda^{1/2}$ at the minimum of the potential. To expand around this minimum we set $\phi=S(x)+\tilde{\phi}$ where $\tilde{\phi}$ is the quantum fluctuation. The fluctuation potential then takes the form $V(\tilde{\phi})=-\mu^4/4\lambda+\mu^2\tilde{\phi}^2+\lambda^{1/2}\mu\tilde{\phi}^3+\lambda \tilde{\phi}^4/4$. Consequently, the initial $\phi\rightarrow -\phi$ symmetry of $V(\phi)$ is lost with $V(\tilde{\phi})$ not having any $ \tilde{\phi} \rightarrow -\tilde{\phi}$ symmetry.

This same pattern is repeated for fermions  and is readily seen in the paradigm for dynamical symmetry breaking, namely the chirally-symmetric Nambu-Jona-Lasinio model (NJL) \cite{Nambu1961}. As formulated  in flat spacetime the model consists of a fermion coupled to a four-fermion interaction with no intrinsic fermion mass term and action
\begin{eqnarray}
I_{\rm NJL}&=&\int d^4x \left[i\bar{\psi}\gamma^{\mu}\partial_{\mu}\psi-\frac{g}{2}[\bar{\psi}\psi]^2-\frac{g}{2}[\bar{\psi}i\gamma_5\psi]^2\right].
\label{6.1x}
\end{eqnarray}
One introduces a mass as a trial parameter and rewrites $I_{\rm NJL}=I_{MF}+I_{RI}$  in terms of  mean field and residual interaction components of the form
\begin{eqnarray}
I_{MF}&=&\int d^4x \left[i \bar{\psi}\gamma^{\mu}\partial_{\mu}\psi-m\bar{\psi}\psi +\frac{m^2}{2g}\right],\qquad 
I_{RI}=\int d^4x \left[-\frac{g}{2}\left(\bar{\psi}\psi-\frac{m}{g}\right)^2-\frac{g}{2}\left(\bar{\psi}i\gamma_5\psi\right)^2\right].
\label{6.2x}
\end{eqnarray}
(This is equivalent to replacing $\tfrac{1}{2}[\bar{\psi}\psi]^2$ by $\bar{\psi}\psi\langle \Omega|\bar{\psi}\psi|\Omega\rangle$.)
In the decomposition of the $I_{\rm NJL}$ action into mean-field and residual interaction components we note that neither of these two components is separately chiral symmetric, with it being only their sum that is. The residual interaction thus restores the chiral symmetry that is lost in the mean-field sector, and in doing so it generates massless pseudoscalar Goldstone bosons  \cite{Nambu1961} (as required by the axial-vector Ward identity) and massive scalar Higgs bosons (see \cite{Nambu1961} and e.g.  \cite{Mannheim2017}) as dynamical bound states. 

In the Hartree-Fock approximation one sets 
\begin{align}
&\langle \Omega|\left[\bar{\psi}\psi-\frac{m}{g}\right]^2|\Omega\rangle=\langle \Omega|\left[\bar{\psi}\psi-\frac{m}{g}\right]|\Omega\rangle^2=0,\quad m= g\langle \Omega|\bar{\psi}\psi|\Omega\rangle,\quad
\langle \Omega|\left[\bar{\psi}i\gamma_5\psi\right]^2|\Omega\rangle=\langle \Omega|\left[\bar{\psi}i\gamma_5\psi\right]|\Omega\rangle^2=0,
\label{6.3x}
\end{align}
as satisfied by a self-consistent $M$ according to
\begin{eqnarray}
-\frac{M\Lambda^2}{4\pi^2}+\frac{M^3}{4\pi^2}{\rm ln}\left(\frac{\Lambda^2}{M^2}\right)=\frac{M}{g},
\label{6.4x}
\end{eqnarray}
where $\Lambda$ is a large momentum cutoff. 
In this approximation  the vacuum energy is given by the non-chiral-invariant mean-field sector alone since $\langle \Omega|I_{RI}|\Omega\rangle=0$, and is of the form \cite{Eguchi1974,Mannheim1976}
\begin{eqnarray}
\tilde{\epsilon}(m)&=&\epsilon(m)-\frac{m^2}{2g}=i\int \frac{d^4p}{(2\pi)^4}{\rm Tr~ln}\left[\slashed{p}-m+i\epsilon\right]-i\int \frac{d^4p}{(2\pi)^4}{\rm Tr~ln}\left[\slashed{p}+i\epsilon\right]
-\frac{m^2}{g}
\nonumber\\
&=&
\frac{m^4}{16\pi^2}{\rm ln}\left(\frac{\Lambda^2}{m^2}\right)-\frac{m^2M^2}{8\pi^2}{\rm ln}\left(\frac{\Lambda^2}{M^2}\right)+\frac{m^4}{32\pi^2},
\label{6.5x}
\end{eqnarray}
where the trace is taken over an occupied negative energy Dirac sea composed of plane waves. The vacuum energy given in (\ref{6.5x}) has the form of a dynamically generated double-well potential with a minimum at $m=M$. The effect of the residual interaction is to then radiatively dress the mean-field sector while not changing the Hilbert space built on the vacuum in which $\langle \Omega|\bar{\psi}\psi|\Omega\rangle$ is non-zero. For our purposes here we note that even though we have generated the fermion mass dynamically through the choice of vacuum,  as can be seen from the structure of the mean-field $I_{MF}$, the orbits associated with the mean-field sector are same as those associated with a fermion with a mechanical mass, i.e., the standard massive particle geodesics that are associated with the test particle action $I_T=-m\int ds$ with a constant mass parameter $m$.

If we take $m$ to be a function of the coordinates we have to evaluate ${\rm Tr~ln}\left[i\slashed{\partial}-m(x)+i\epsilon\right]$ -${\rm Tr~ln}\left[i\slashed{\partial}+i\epsilon\right]$ (still for the moment in flat spacetime), where now the trace is taken over an occupied negative energy Dirac sea composed of distorted waves. And on doing so obtain an effective action whose leading term is of the logarithmically divergent  form \cite{Eguchi1974,Mannheim1976} 
\begin{eqnarray}
I_{\rm EFF}=\int \frac{d^4x}{8\pi^2}{\rm ln}\left(\frac{\Lambda^2}{M^2}\right)\bigg[
\frac{1}{2}\eta_{\mu\nu}\partial^{\mu}m(x)\partial^{\nu}m(x)
+m^2(x)M^2-\frac{1}{2}m^4(x)\bigg],
\label{6.6x}
\end{eqnarray}
as written here with metric signature $g_{00}>0$. In (\ref{6.6x}) we identify $m(x)$ as $m(x)=g\langle C|\bar{\psi}\psi |C\rangle$, where $|C\rangle$ is a coherent state.  As constructed, on its own the state $|C\rangle$ is not stable. To stabilize it we need to introduce a positive frequency fermion state. Thus we add on to (\ref{6.6x}) a positive frequency one-particle action term 
\begin{eqnarray}
I(1)=\int d^4x \left[i \bar{\psi}(1)\gamma^{\mu}\partial_{\mu}\psi(1)-m(x)\bar{\psi}(1)\psi(1) \right],
\label{6.7y}
\end{eqnarray}
with the stationary minimum then being the state $u_0(x)$ introduced above.
With $m(x)$ only varying within the fermion and with it taking a constant value $m$ outside, the fermion propagates with mass parameter $m(x)=m$ (i.e., propagation is not affected by internal structure, just as in standard Einstein gravity). Trajectories associated with (\ref{6.7y}) are thus given by setting $m(x)=m$ in (\ref{6.7y}), to thus be standard flat space massive particle geodesics.

Finally, if we evaluate ${\rm Tr~ln}\left[i\gamma^aV^{\mu}_a(\partial_{\mu}+\Gamma_{\mu})-m(x)+i\epsilon\right]$ -${\rm Tr~ln}\left[i\gamma^aV^{\mu}_a(\partial_{\mu}+\Gamma_{\mu})+i\epsilon\right]$ 
in a background curved space geometry we obtain \cite{'tHooft2010a}
\begin{eqnarray}
I_{\rm EFF}=\int \frac{d^4x}{8\pi^2}(-g)^{1/2}{\rm ln}\left(\frac{\Lambda^2}{M^2}\right)\bigg[
\frac{1}{2}g_{\mu\nu}\nabla^{\mu}m(x)\nabla^{\nu}m(x)+\frac{1}{12}m^2(x)R^{\alpha}_{\phantom{\alpha}\alpha}
+m^2(x)M^2-\frac{1}{2}m^4(x)\bigg].
\label{6.8y}
\end{eqnarray}
With (\ref{6.7y}) being replaced by
\begin{eqnarray}
I(1)=\int d^4x (-g)^{1/2}\left[i \bar{\psi}(1)[i\gamma^aV^{\mu}_a(\partial_{\mu}+\Gamma_{\mu})\psi(1)-m(x)\bar{\psi}(1)\psi(1) \right], 
\label{6.9y}
\end{eqnarray}
and with $m(x)$ being constant outside the source, 
trajectories associated with (\ref{6.9y}) (i.e.,  associated with the positive energy modes that are observed in astrophysics) are thus standard curved space massive particle geodesics.

We recognize (\ref{6.8y}) as being a dynamical symmetry generalization of the (generic) conformally coupled scalar field action given in (\ref{1.6y}), with the c-number $m(x)=g\langle C|\bar{\psi}\psi|C\rangle$ replacing $S(x)$ (i.e., $m(x)$ is not a quantized field that has a canonical conjugate with which it would not commute).  Consequently, $S(x)$ is not an elementary field that appears in a fundamental Lagrangian (i.e., a field that would have a canonical conjugate with which it would not commute).  As can be seen from the middle panel in Fig. \ref{bag}, $m(x)$ only varies in the same region as the region where the bound positive energy fermion wave function $u_0(x)$ varies, i.e., solely within the extended fermion source. There is thus no macroscopic scalar field that varies outside the source.  Moreover, since the $R^{\alpha}_{\phantom{\alpha}\alpha}$ term that is generated in (\ref{6.8y}) is associated with the $g_{\mu\nu}\partial^{\mu}m(x)\partial^{\nu}m(x)$ term (only the combination of the two of them is locally conformal invariant), it also is only present in the region where $m(x)$ varies, i.e., solely within the extended source. As we see, through one and the same dynamical symmetry breaking mechanism we generate masses for fermions through $\langle \Omega|\bar{\psi}\psi |\Omega \rangle$ and localize them through $\langle C|\bar{\psi}\psi |C\rangle$. 

While (\ref{6.4x}), (\ref{6.5x}), (\ref{6.6x}) and  and (\ref{6.8y}) involve the cutoff, on using a renormalization group fixed point analysis with subtraction parameter $\mu^2$, in \cite{Mannheim1978,Mannheim2017} it was shown that via critical scaling with anomalous dimensions we could make $\tilde{\epsilon}(m)$ and $I_{\rm EFF}$ be completely finite \cite{footnoteF}. Specifically, it was shown that if the dimension $d$ of $\theta=\bar{\psi}\psi$ is reduced from its canonical value of $d=3$ to $d=2$ (i.e., $\theta$ acquires a dimension $d=3+\gamma_{\theta}$ where $\gamma_{\theta}=-1$) then the vacuum spontaneously breaks and one obtains dynamical Goldstone and Higgs bosons. With the dimension of $(\bar{\psi}\psi)^2$ being reduced from six to four the four-fermion theory becomes renormalizable, the coupling parameter $g$ in $I_{\rm NJL}$ action becomes dimensionless, and the theory is now scale invariant, just as required for a matter sector that is to couple to conformal gravity. 

At $\gamma_{\theta}=-1$ (\ref{6.8y}) is replaced by \cite{Mannheim1978,Mannheim2017}
\begin{align}
I_{\rm EFF}=\int \frac{d^4x}{8\pi^2}(-g)^{1/2}\bigg{[}
-\frac{m^2(x)\mu^2}{16\pi^2}\left({\rm ln}\left(\frac{m^2(x)}{M^2}\right)-1\right)
+\frac{3\mu}{256\pi m(x)}\left(g_{\kappa\nu}\nabla^{\kappa}m(x)\nabla^{\nu}m(x)+\frac{1}{6}m^2(x)R^{\alpha}_{\phantom{\alpha}\alpha}\right)\bigg{]},
\label{6.13y}
\end{align}
with the already curved space, non-conformal invariant, mean-field induced (\ref{6.9y}) remaining as is.

This localization mechanism generates cosmological inhomogeneities in conformal gravity, and allows them to propagate in an exterior geometry in which both $\nabla_{\mu}m$ and $R^{\alpha}_{\phantom{\alpha}\alpha}$ are zero. In such regions in which $f(r)$ is zero one thus has $B(r)=B_0(r)+B^*(r)$, while orbital velocities are given by the standard $(r^2/c^2)(d\phi/dt)^2=rB^{\prime}/2$ that was used in \cite{Mannheim2012,Mannheim2011,O'Brien2012}. Thus unlike in the case with a long range, massless scalar field, when mass is generated  microscopically via dynamical symmetry breaking  orbits are standard. 

\begin{acknowledgments}
The author wishes to thank Dr. Y. Brihaye, Dr. M. P. Hobson and  Dr. A. N. Lasenby for some helpful comments.
\end{acknowledgments}

\appendix
\numberwithin{equation}{section}
\setcounter{equation}{0}

\section{Connection with Cosmology}
\label{S1C}

In a Robertson-Walker geometry the comoving time line element is given by  
\begin{eqnarray}
ds^2=c^2dt^2-\frac{a^2(t)}{(1+k\rho^2/4)^2}[d\rho^2+\rho^2d\theta^2+\rho^2\sin^2\theta d\phi^2]
\label{A.10zy} 
\end{eqnarray}
as written in spatially isotropic coordinates, where $a(t)$ is the expansion radius and $k$ is the spatial three-curvature \cite{footnoteA}. Since the Robertson-Walker metric is  conformal to flat,  the conformal $W_{\mu\nu}$ tensor vanishes identically. We take the cosmological energy-momentum tensor to consist of a conformally coupled scalar field $\sigma(x)$ and a $T_0^{\mu \nu}=(\rho_0+p_0)U^{\mu}U^{\nu}+p_0g^{\mu\nu}$ perfect fluid that does not couple to the scalar field (as noted in Sec. \ref{S1}, cosmological or grandunified scales are not associated with the mass generation of the standard model elementary particles that make up the $T_0^{\mu \nu}$ perfect fluid). In analog to (\ref{1.8y}) and (\ref{1.9y})  the scalar field wave equation and the full energy-momentum tensor are given by
\begin{eqnarray}
&&\nabla_{\mu}\nabla^{\mu}\sigma+\frac{1}{6}\sigma R^\mu_{\phantom{\mu}\mu}
-4\lambda_0 \sigma^3 =0,
\nonumber\\
&&T^{\mu \nu}=(\rho_0+p_0)U^{\mu}U^{\nu}+p_0g^{\mu\nu} +\frac{2}{3}\nabla^{\mu}\sigma\nabla^{\nu} \sigma
-\frac{1}{6}g^{\mu\nu}\nabla_{\alpha}\sigma\nabla^{\alpha}\sigma
-\frac{1}{3}\sigma\nabla^{\mu}\nabla^{\nu}\sigma
\nonumber\\             
&&+\frac{1}{3}g^{\mu\nu}\sigma\nabla_{\alpha}\nabla^{\alpha}\sigma                           
-\frac{1}{6}\sigma^2\left(R^{\mu\nu}
-\frac{1}{2}g^{\mu\nu}R^\alpha_{\phantom{\alpha}\alpha}\right)-g^{\mu\nu}\lambda_0 \sigma^4, 
\label{A.11zy}
\end{eqnarray}                                 
where the energy density $\rho_0$ and the pressure $p_0$ are functions of $t$, and $U_{\mu}$ is a timelike fluid four-velocity that obeys $g_{\mu\nu}U^{\mu}U^{\nu}=-1$ (in the notation of \cite{Weinberg1972} $g_{00}$ is negative).   With there being no cross term between the  $\sigma$ sector and the fluid sector,  the $\sigma$ sector contribution and the perfect fluid contribution to $T^{\mu\nu}$ are both separately covariantly conserved, with the $\sigma$ sector containing a cosmological constant term $-g^{\mu\nu}\lambda_0 \sigma^4$. The full $T^{\mu\nu}$ must vanish identically since $W^{\mu\nu}$ does, and it is able to  do so non-trivially \cite{Mannheim2006}. With $\sigma(x)$ taking a constant value $\sigma_0$,  with the fluid being taken to be a radiation fluid prior to the mass generation epoch so that $\rho_0=3p_0=A/a^4(t)$ (where $A$ is a constant), and with $k$ being shown to be negative in the conformal case \cite{Mannheim2006}, the exact equations and solution for the expansion radius $a(t)$ in the vanishing $T^{\mu\nu}$ case are given as \cite{Mannheim2006}
\begin{eqnarray}
&&\frac{A}{a^4}+\lambda_0\sigma_0^4+\frac{1}{2a^2c^2}\sigma_0^2(\dot{a}^2+kc^2)=0,\quad \dot{a}^2+a\ddot{a}+kc^2+4c^2\lambda_0\sigma_0^2a^2=0,
\nonumber\\
&&a^2(t)= -\frac{k(\beta-1)}{2\alpha}
-\frac{k\beta{\rm sinh}^2 (\alpha^{1/2} ct)}{\alpha},
\label{A.12zy}
\end{eqnarray}
where the parameters $\alpha$ and $\beta$ are defined as 
\begin{eqnarray}
 \alpha=-2\lambda_0 \sigma_0^2=K_0,\quad
\beta=\left(1- \frac{16A\lambda_0}{k^2}\right)^{1/2},
\label{A.13zy}
\end{eqnarray}
and where we have  introduced $K_0=-2\lambda_0\sigma_0^2$, a parameter that serves as a de Sitter scale. 
With this model a non-fine-tuned, dark matter free, fit to the accelerating universe data is obtained \cite{Mannheim2006}, with $k$  indeed being  found to be negative. 

\section{Writing a Robertson-Walker Geometry in Static Coordinates}
\label{S1D}

In order to be able to make contact between cosmology and galaxies, this being of relevance to our analysis above, it is very instructive to rewrite the Robertson-Walker line element in a static, spherically symmetric coordinate system. That we are able to do this in principle is because in a geometry that is homogeneous and isotropic any point in the spacetime can be taken as the origin of coordinates. Thus each time we analyze the rotation curve of any given galaxy we take the center of that particular  galaxy to be the origin of coordinates.

To establish the connection that we need we follow \cite{Mannheim1989} and make the coordinate transformation 
\begin{eqnarray}
\rho&=&\frac{8+4\gamma_0 r-8(1+\gamma_0r-K_0r^2)^{1/2}}{(\gamma_0^2+4K_0)r},\quad
r=\frac{16\rho}{16-8\gamma_0\rho+\gamma_0^2\rho^2+4K_0\rho^2},
\nonumber\\
&&\frac{\rho}{1+k\rho^2/4}=\frac{\rho}{1-\gamma_0\rho^2/16-K_0\rho^2/4}=\frac{r}{(1+\gamma_0r-K_0r^2)^{1/2}},
\label{A.14zy}
\end{eqnarray}
with (\ref{A.10zy}) then taking the form
\begin{eqnarray}
ds^2=c^2dt^2-a^2(t)\bigg{(}\frac{dr^2}{(1+\gamma_0r-K_0r^2)^2}+\frac{r^2}{1+\gamma_0r-K_0r^2}(d\theta^2+\sin^2\theta d\phi^2)\bigg{)},
\label{A.15zy} 
\end{eqnarray}
where $-\gamma_0^2/4-K_0=k$. With the de Sitter $K_0$ being positive and with $\gamma_0$ being real and non-zero, it follows that the only allowed value of $k$ for which we can in fact make this transformation at all is negative, just as noted above.

Introducing the conformal time $d\tau= dt/a(t)$ enables us to rewrite (\ref{A.15zy}) in the form
\begin{eqnarray}
ds^2=a^2(\tau)\bigg{[}c^2d\tau^2-\frac{dr^2}{(1+\gamma_0r-K_0r^2)^2}-\frac{r^2}{1+\gamma_0r-K_0r^2}(d\theta^2+\sin^2\theta d\phi^2)\bigg{]},
\label{A.16zy} 
\end{eqnarray}
with (\ref{A.16zy}) still describing a Robertson-Walker geometry.
Noting now that the Weyl tensor not only vanishes in a Robertson-Walker geometry but also in a conformally transformed one, we now conformally transform (\ref{A.16zy}) by multiplying the line element by $\Omega^2(r,\tau)$ where $\Omega(r,\tau)=(1+\gamma_0r-K_0r^2)^{1/2}/a(\tau)$, with (\ref{A.16zy}) then taking the form
\begin{eqnarray}
ds^2=(1+\gamma_0r-K_0r^2)c^2d\tau^2-\frac{dr^2}{(1+\gamma_0r-K_0r^2)}-r^2(d\theta^2+\sin^2\theta d\phi^2).
\label{A.17zy} 
\end{eqnarray}
Since this line element is now of the same generic form as the  static, spherically symmetric line element $ds^2=B(r)dt^2-dr^2/B(r)-r^2d\theta^2-r^2\sin^2\theta d\phi^2$ that we met in (\ref{2.6x}) (on replacing $\tau$ by $t$), we see that a Robertson-Walker geometry is conformally and coordinate equivalent to a static, spherically symmetric geometry, viz. precisely the geometry needed for study of galactic rotation curves, with cosmology precisely providing the $B_0(r)=1+\gamma_0r-K_0r^2$ contribution to $B(r)$ that we identified in Sec. \ref{S2}. 

In making the conformal transformation the time-dependent $\rho_0=A/a^4(\tau)$ transforms into the static $\Omega^{-4}\rho_0=A/(1+\gamma_0r-K_0r^2)^2$, while the time-independent $\sigma_0$ transforms into the time-dependent $\Omega^{-1}\sigma_0=\sigma_0a(\tau)/(1+\gamma_0r-K_0r^2)^{1/2}$. To see how these dynamical conditions are obeyed (something that had not been considered in the purely kinematic study given in \cite{Mannheim1989} and \cite{Mannheim2006}), we note that for the line element $ds^2=B(r)dt^2-dr^2/B(r)-r^2d\theta^2-r^2\sin^2\theta d\phi^2$, the covariant conservation condition for a perfect fluid that only depends on $r$ is of the form $B^{\prime}/B=-2p_0^{\prime}/(\rho_0+p_0)$. (This relation actually holds for the general $ds^2=B(r)dt^2-A(r)dr^2-r^2d\theta^2-r^2\sin^2\theta d\phi^2$.) Thus for a fluid that obeys $\rho_0=3p_0$, $\rho_0$ must behave as $1/B^2$, i.e., precisely as $1/(1+\gamma_0r-K_0r^2)^2$ just as we had found. For the scalar field, we find that the time-dependent $\sigma(r,\tau) =\sigma_0a(\tau)/(1+\gamma_0r-K_0r^2)^{1/2}$ will satisfy the scalar field wave equation given in (\ref{A.11zy}) as evaluated with the (\ref{A.17zy}) metric provided $a(\tau)$ obeys 
\begin{eqnarray}
\frac{d^2a(\tau)}{d\tau^2}+kc^2a(\tau)+4\lambda_0\sigma_0^2c^2a^3(\tau)=0.
\label{A.18zy}
\end{eqnarray}
On converting the second equation in (\ref{A.12zy}) from comoving time to conformal time we recover (\ref{A.18zy}) just as required.

With particles being massless in the radiation era they propagate along null geodesics, with these geodesics  thus being sensitive to both the $\gamma_0$ and $K_0$ terms in (\ref{A.17zy}). This dependence will persist after mass generation, and so both of these terms will also contribute to the trajectories followed by massive particles as well. The $\gamma_0 r$ term is particularly significant for conformal gravity rotation curve fits, and is found numerically to be one of the largest contributors, with it overwhelming the contribution of any given star to the galactic potential by a factor of order $10^{11}$. Thus as noted in \cite{Mannheim2006}, cosmology is central to the systematics of galactic rotation curves.


\begin{thebibliography}{99}

\bibitem{Brans1961} C. H. Brans and R. H. Dicke, \href{https://doi.org/10.1103/PhysRev.124.925} {Phys. Rev. \textbf{124}, 925 (1961).}

\bibitem{Weinberg1972} S. Weinberg, {\it Gravitation and Cosmology:
Principles  and Applications of the General Theory of Relativity} (Wiley, New York, 1972).

\bibitem{Brihaye2009} Y. Brihaye and Y. Verbin, \href{https://doi.org/10.1103/PhysRevD.80.124048}{Phys. Rev. D \textbf{80}, 124048 (2009).}

\bibitem{Horne2016} K. Horne, \href{https://doi.org/10.1093/mnras/stw506}{ Mon. Not. R. Astron. Soc. \textbf{458}, 4122 (2016).}

\bibitem{Hobson2021} M. P. Hobson and A. N. Lasenby, \href{https://doi.org/10.1103/PhysRevD.104.064014}{Phys. Rev. D \textbf{104}, 064014 (2021).}

 \bibitem{Mannheim2011} P. D. Mannheim and J. G. O'Brien,  \href{https://doi.org/10.1103/PhysRevLett.106.121101}{Phys. Rev. Lett. \textbf{106}, 121101 (2011).}


 \bibitem{Mannheim2012} P. D. Mannheim and J. G. O'Brien, \href{https://doi.org/10.1103/PhysRevD.85.124020}{Phys. Rev. D \textbf{85}, 124020 (2012).}

 
\bibitem{O'Brien2012} J. G. O'Brien and P. D. Mannheim, \href{https://doi.org/10.1111/j.1365-2966.2011.20386.x}{ Mon. Not. R. Astron. Soc. \textbf{421,} 1273 (2012).}

\bibitem{Mannheim1994} P. D. Mannheim and D. Kazanas, \href{https://doi.org/10.1007/BF02105226}{Gen. Rel. Gravit. \textbf{26}, 337 (1994).}


\bibitem{Mannheim1994a} P. D. Mannheim, {\it Four dimensional conformal gravity, confinement, and 
galactic rotation curves}, in Proceedings of ``PASCOS 94", the Fourth 
International Symposium on Particles, Strings and Cosmology, Syracuse, New York,        
May 1994. Edited by K. C. Wali, World Scientific Press, Singapore (1995). \href{https://arxiv.org/abs/gr-qc/9407010}
{(arXiv:gr-qc/9407010.)}


\bibitem{Mannheim2006} P. D. Mannheim,  \href{https://doi.org/10.1016/j.ppnp.2005.08.001}{Prog. Part. Nucl. Phys. \textbf{56}, 340 (2006).}

\bibitem{Mannheim2017} P. D. Mannheim,  \href{http://dx.doi.org/10.1016/j.ppnp.2017.02.001}{Prog. Part. Nucl. Phys. \textbf{94}, 1250 (2017).}

\bibitem{Mannheim1989} P. D. Mannheim and D. Kazanas, \href{https://doi.org/10.1086/167623}{Astrophys. J. \textbf{342}, 635 (1989).}


\bibitem{Riegert1984b} R. J. Riegert, \href{https://doi.org/10.1103/PhysRevLett.53.315}{Phys. Rev. Lett. {\bf 53}, 315 (1984).}

\bibitem{footnoteC} As well as this particular solution, we note in passing that in \cite{Brihaye2009} some other solutions were also considered, solutions that provide insight into black holes and solitons in a  conformal gravity context.



\bibitem{Mannheim1991} P. D. Mannheim and D. Kazanas, \href{https://doi.org/10.1103/PhysRevD.44.417} {Phys. Rev. D \textbf{44}, 417 (1991).}


 \bibitem{Mannheim1993} P. D. Mannheim,  \href{https://doi.org/10.1007/BF00756938}{Gen. Rel. Gravit. \textbf{25}, 697 (1993).}
 
  \bibitem{footnoteE} Equation (\ref{4.5y}) corrects equation (A41) of \cite{Mannheim2012}.

 
 \bibitem{Mannheim2021}  P. D. Mannheim, \textit{Critique of the use of geodesics in astrophysics and cosmology}, \href{https://arxiv.org/abs/2105.08556}{arXiv:2105.08556 [gr-qc].}

 
 



\bibitem{Dashen1974} R. F. Dashen, B. Hasslacher and A. Neveu, \href{https://doi.org/10.1103/PhysRevD.10.4130}{Phys. Rev. D 
\textbf{10}, 4130 (1974).}

\bibitem{Bardeen1975} W. A. Bardeen, M. S. Chanowitz, S. D. Drell, M. Weinstein, T.-M. Yan, \href{https://doi.org/10.1103/PhysRevD.11.1094}{Phys. Rev. D  \textbf{11}, 1094 (1975).}

\bibitem{Dashen1975} R. F. Dashen, B. Hasslacher and A. Neveu, \href{https://doi.org/10.1103/PhysRevD.12.2443}{Phys. Rev. D \textbf{12}, 2443 (1975).}

\bibitem{Mannheim1976}  P. D. Mannheim, \href{https://doi.org/10.1103/PhysRevD.14.2072}{Phys. Rev. D \textbf{14}, 2072 (1976).}



\bibitem{Nambu1961} Y. Nambu and G. Jona-Lasinio, \href{https://doi.org/10.1103/PhysRev.122.345}{Phys. Rev. \textbf{122}, 345 (1961).}



\bibitem{Eguchi1974}  T. Eguchi and  H. Sugawara, \href{https://doi.org/10.1103/PhysRevD.10.4257}{Phys. Rev. D \textbf{10}, 4257 (1974).}

\bibitem{'tHooft2010a} G. 't Hooft, \textit{Probing the small distance structure of canonical quantum gravity using the conformal group}, September 2010,  \href{https://arxiv.org/abs/1009.0669}{arXiv:1009.0669 [gr-qc].}


\bibitem{Mannheim1978}  P. D. Mannheim, \href{https://doi.org/10.1016/0550-3213(78)90025-1}{Nucl. Phys. B \textbf{143}, 285 (1978).}

\bibitem{footnoteF} To implement critical scaling we augment the four-fermion theory with an electrodynamics that is at a renormalization group fixed point. At the fixed point the Green's function for the insertion  of $\bar{\psi}\psi$ into the inverse fermion propagator then scales as $\tilde{\Gamma}_{\rm S}(p,p+q,q,m=0)=\left[(-p^2/\mu^2)(-(p+q)^2)/\mu^2)\right]^{\gamma_{\theta}/4}$. Dressing the point four-fermion vertices (viz. vertices with $\tilde{\Gamma}_{\rm S}(p,p+q,q,m=0)=1$) with this $\tilde{\Gamma}_{\rm S}(p,p+q,q,m=0)$ then leads to (\ref{6.8y}) when $\gamma_{\theta}=-1$ 
\cite{Mannheim1978,Mannheim2017}.


\bibitem{footnoteA} In terms of the coordinate system in which $ds^2=c^2dt^2-a^2(t)[dr^2/(1-kr^2)+r^2d\theta^2+r^2\sin^2\theta d\phi^2]$, $r$ and $\rho$ are related by $r=\rho/(1+k\rho^2/4)$, $\rho=2[1-(1-kr^2)^{1/2}]/kr$.






\end{thebibliography}
\end{document}